\newif\ifonecol
\onecolfalse 

\ifonecol 
\documentclass[11pt, journal, draftclsnofoot, onecolumn]{IEEEtran}
\else 
\documentclass[11pt, letterpaper]{IEEEtran}
\fi

\usepackage{times,amsmath,epsfig,algorithm,amsfonts,amssymb,theorem,subfigure,citesort,float,dblfloatfix,comment,framed}
\usepackage[noend]{algorithmic}

\newcommand{\qed}{\nobreak \ifvmode \relax \else
      \ifdim\lastskip<1.5em \hskip-\lastskip
      \hskip1.5em plus0em minus0.5em \fi \nobreak
      \vrule height0.75em width0.5em depth0.25em\fi}

\floatstyle{boxed}
\newfloat{inset}{thp}{lop}
\floatname{inset}{Sidebar}

\newcounter{bx}

\ifonecol 
\title{\Huge{Secure Biometrics}\\ \LARGE{Concepts, Authentication Architectures \& Challenges}}
\else 
\title{\huge{Secure Biometrics:\\ Concepts, Authentication Architectures \& Challenges}}
\fi
\author{
{Shantanu Rane, Ye Wang, Stark C. Draper, and Prakash Ishwar}%
\vspace{1.2mm}\\
}
\begin{document}
\maketitle

\IEEEPARstart{B}{iometrics} are an important and widely used class of methods for
identity verification and access control.  Biometrics are attractive
because they are inherent properties of an individual. They need not
be remembered like passwords, and are not easily lost or forged like
identifying documents. At the same time, biometrics are fundamentally
noisy and irreplaceable. There are always slight variations among
the measurements of a given biometric, and, unlike passwords or
identification numbers, biometrics are derived from physical
characteristics that cannot easily be changed. The
proliferation of biometric usage raises critical privacy and security concerns that, 
due to the noisy nature of biometrics, cannot be addressed using standard
cryptographic methods. In this article we present an overview of ``secure biometrics'', also referred to as
``biometric template protection'', an emerging class of methods that
address these concerns.

The traditional method of accommodating measurement variation among
biometric samples is to store the enrollment sample on the device, and
to match it against a probe provided by the individual being
authenticated.  Consequently, much effort has been invested in the
development of pattern recognition algorithms for biometric matching
that can accommodate these variations.  Unfortunately, this approach
has a serious flaw: An attacker who steals or hacks into the device
gains access to the enrollment biometric.  In conventional
password-based systems, this type of problem can be mitigated by
storing a non-invertible cryptographic hash of the password rather
than the password itself. However, cryptographic hashes are extremely
sensitive to noise, and thus incompatible with the inherent variability
of biometric measurements. Therefore, the above approach used for 
securing passwords is
ill-suited to biometric security.

The loss of an enrollment biometric to an attacker is a security
hazard because it may allow the attacker to gain unauthorized access
to facilities, sensitive documents, and the finances of the victim.
Further, since a biometric signal is tied to unique physical
characteristics and the identity of an individual, a leaked biometric
can result in a significant loss of privacy.  In this article, we
refer to a {\em security} breach as an event wherein an attacker
successfully accesses a device.  We refer to a {\em privacy} breach as
an event wherein an attacker partially, or completely, determines the
victim's biometric.  Security and privacy breaches represent distinct
kinds of attacks.  Significantly, the occurrence of one does not
necessarily imply the occurrence of the other.

Addressing these challenges demands new approaches to the
design and deployment of biometric systems. We refer to these as ``secure
biometric'' systems. Research into secure biometrics has drawn on
advances in the fields of signal processing~\cite{nandakumar07tifs,nagar08icpr,uludag04workbiom,sutcu07tifs,clancy03acmbma,yang05icassp}, 
error correction coding~\cite{juels99fuzzy,juels06jdcc,draper07icassp,sutcu08isit,wang09wifs},
information theory~\cite{lai11tifs1,lai11tifs2,ignatenko09tifs,ignatenko2012fnt} and
cryptography~\cite{blanton11esorics,barni10mmsec,sadeghi09icisc}.
Four main architectures dominate: fuzzy commitment, secure sketch, secure multiparty 
computation, and cancelable biometrics.  The first two architectures, 
fuzzy commitment and secure sketch provide
information-theoretic guarantees for security and privacy, using error 
correcting codes or signal embeddings. The third architecture 
attempts to securely determine the distance between
enrollment and probe biometrics, using computationally secure 
cryptographic tools such as garbled circuits and homomorphic 
encryption.  The final architecture, cancelable biometrics, involves
distorting the biometric signal at enrollment with a secret
user-specific transformation, and storing the distorted biometric at
the access control device. 

It is the aim of this article to provide a tutorial overview of these
architectures. To see the key commonalities and differences among the
architectures, it is useful to first consider a generalized framework for secure
biometrics, composed of biometric encoding and decision-making stages.
For this framework, we can precisely characterize performance in 
terms of metrics for accuracy, security and privacy. Furthermore, it
is vital to understand the statistical properties and constraints that must
be imposed on biometric feature extraction algorithms, in order to make
them viable in a secure biometric system. Having presented the general
framework, and specified constraints on feature extraction, we can then
cast the four architectures listed above as specific realizations of the 
generalized framework, allowing the reader to compare and contrast them 
with ease. The discussion of single biometric access control systems naturally
leads to questions about multi-system deployment, i.e., the situation in which 
a single user has enrolled his or her biometric on multiple devices.
An analysis of the multi-system case reveals interesting privacy-security 
tradeoffs that have been only minimally analyzed in the
literature. One of our goals is to highlight interesting open problems
related to multi-system deployment in particular and secure biometrics
in general, and to spur new research in the field.
\section{A Unified Secure Biometrics Framework}
\label{sec:framework}

Secure biometrics may be viewed as a problem of designing a
suitable \emph{encoding} procedure for transforming an
enrollment biometric signal into data to be stored on the
authentication device, and of designing a matching \emph{decoding}
procedure for combining the probe biometric signal with the
stored data to generate an \emph{authentication decision}.  This
system is depicted in Figure~\ref{fig:highlevelBD}.  Any analysis of the
privacy and security tradeoffs in secure biometrics must take into
account not only authentication accuracy but also the information
leakage and the possibility of attacking the system when the stored
data and/or keys are compromised. At the outset, note that in authentication, 
a probe biometric is matched against a particular enrollment of one claimed user. This differs 
from identification, in which a probe biometric is matched against each
enrollment in the database to discover the identity associated with the
probe. These are distinct but closely related tasks. For clarity, our
development focuses only on authentication.

\subsection{Biometric Signal Model}

Alice has a biometric --- such as a fingerprint, palmprint, iris,
face, gait, or ECG --- given by nature, that we denote as
$\mathbf{\Lambda}_0$. To enroll at the access control device, Alice
provides a noisy measurement $\mathbf{\Lambda}_E$ of her underlying
biometric $\mathbf{\Lambda}_0$.  From this noisy measurement, a
feature extraction algorithm extracts a feature vector, which we
denote by $\mathbf{A}$. At the time of authentication, Alice provides a
noisy measurement $\mathbf{\Lambda}_P$, from which is extracted a
probe biometric feature vector $\mathbf{B}$. In an attack scenario, an
adversary may provide a biometric signal $\mathbf{\Phi}$, from which
is extracted a biometric feature vector $\mathbf{C}$.
We note here that most theoretical analyses of
secure biometric systems omit the feature extraction step and directly
work with $(\mathbf{A}, \mathbf{B}, \mathbf{C})$ as an abstraction of the biometric signals.
For example, it is convenient to analyze models in which $(\mathbf{A}, \mathbf{B}, \mathbf{C})$ are binary vectors with certain statistical properties.
We will elaborate on the feature extraction process in
an upcoming section, but for the exposition of the system framework, we directly use the feature vectors $\mathbf{A}$ and $\mathbf{B}$ (or $\mathbf{C}$) rather than the underlying biometric signals $\mathbf{\Lambda}_E$, $\mathbf{\Lambda}_P$ and $\mathbf{\Phi}$.

\subsection{Enrollment} 

Consider a general model in which a potentially randomized encoding
function $F(\cdot)$ takes the enrollment feature vector $\mathbf{A}$
as input and outputs stored data $\mathbf{S} \in \mathcal{S}$,
$|\mathcal{S}| < \infty$, which is retained by the access control
device. Optionally, a key vector $\mathbf{K} \in \mathcal{K}$,
$|\mathcal{K}| < \infty$, may also be produced and
  returned to the user, or alternatively, the user can select the key
$\mathbf{K}$ and provide it as another input to
the encoding function.  The arrow in Figure~\ref{fig:generalFramework}
is shown as bi-directional to accommodate these two possibilities,
viz., the system generates a unique key for the user, or the user
selects a key to be applied in the encoding. The
enrollment operation $(\mathbf{S},\mathbf{K}) = F(\mathbf{A})$ (or
$\mathbf{S} = F(\mathbf{A}, \mathbf{K})$ in the case where the key
is user specified) can be described, without loss of generality, by
the conditional distribution $P_{\mathbf{S},\mathbf{K}|\mathbf{A}}$,
which can be further decomposed into various forms and special cases
(e.g., $P_{\mathbf{S}|\mathbf{A}, \mathbf{K}}
P_{\mathbf{K}|\mathbf{A}}$, $P_{\mathbf{S}|\mathbf{A}, \mathbf{K}}
P_{\mathbf{K}}$, $P_{\mathbf{K}|\mathbf{A}, \mathbf{S}}
P_{\mathbf{S}|\mathbf{A}}$, etc.) to specify the exact structure of
how the key and stored data are generated from each other and the
enrollment biometric.  Depending upon the physical realization of
the system, the user may be required to remember the key or carry the
key $\mathbf{K}$, e.g., on a smart card.  Such systems are called {\em
  two-factor} systems because both the ``factors'', namely the
biometric and the key, are needed for authentication and are typically
independent of each other.  In this model, \emph{keyless} (or
single-factor) systems follow in a straightforward way by setting
$\mathbf{K}$ to be null; these do not require separate key storage
(such as a smart card).

\subsection{Authentication} 

As shown in Figure~\ref{fig:generalFramework}, a legitimate user attempts to authenticate by providing a probe
feature vector  $\mathbf{B}$ and the key
$\mathbf{K}$.  An adversary, on the other hand, provides a stolen or
artificially synthesized feature vector $\mathbf{C}$ and a stolen or
artificially synthesized key $\mathbf{J}$. 
Let $(\mathbf{D},\mathbf{L})$ denote the (biometric, key) pair that is provided 
during the authentication step. We write
\[
(\mathbf{D},\mathbf{L}) := \begin{cases}
(\mathbf{B},\mathbf{K}), & \text{if legitimate,}\\
(\mathbf{C},\mathbf{J}), & \text{if adversary.}
\end{cases}
\]
The authentication decision is computed by the binary-valued 
decoding function $g(\mathbf{D},\mathbf{L},\mathbf{S})$. In keyless
systems, the procedure is similar with $\mathbf{K}$, $\mathbf{J}$, and
$\mathbf{L}$ removed from the above description. 
To keep the development simple, we considered only a single 
enrollment $\mathbf{A}$ and a single probe $\mathbf{D}$ above; in practice, using 
multiple biometric measurements during the enrollment or decision phase can 
improve the authentication accuracy~\cite{kelkboom10tsmc}.

\subsection{Performance Measures}

\begin{figure*}
\centering
\ifonecol 
  \subfigure[Authentication System]{
    \includegraphics[width=4.5in]{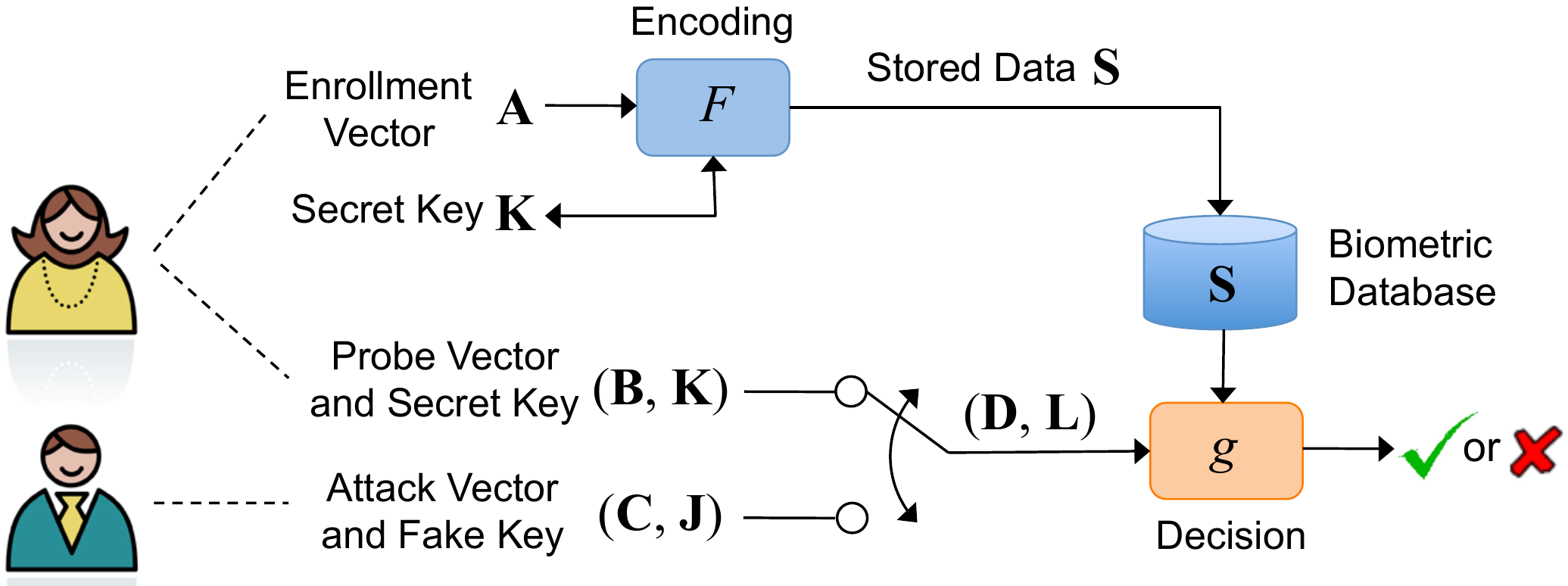}
    \label{fig:generalFramework}}
  \subfigure[ROC Curve]{
    \includegraphics[width=1.7in]{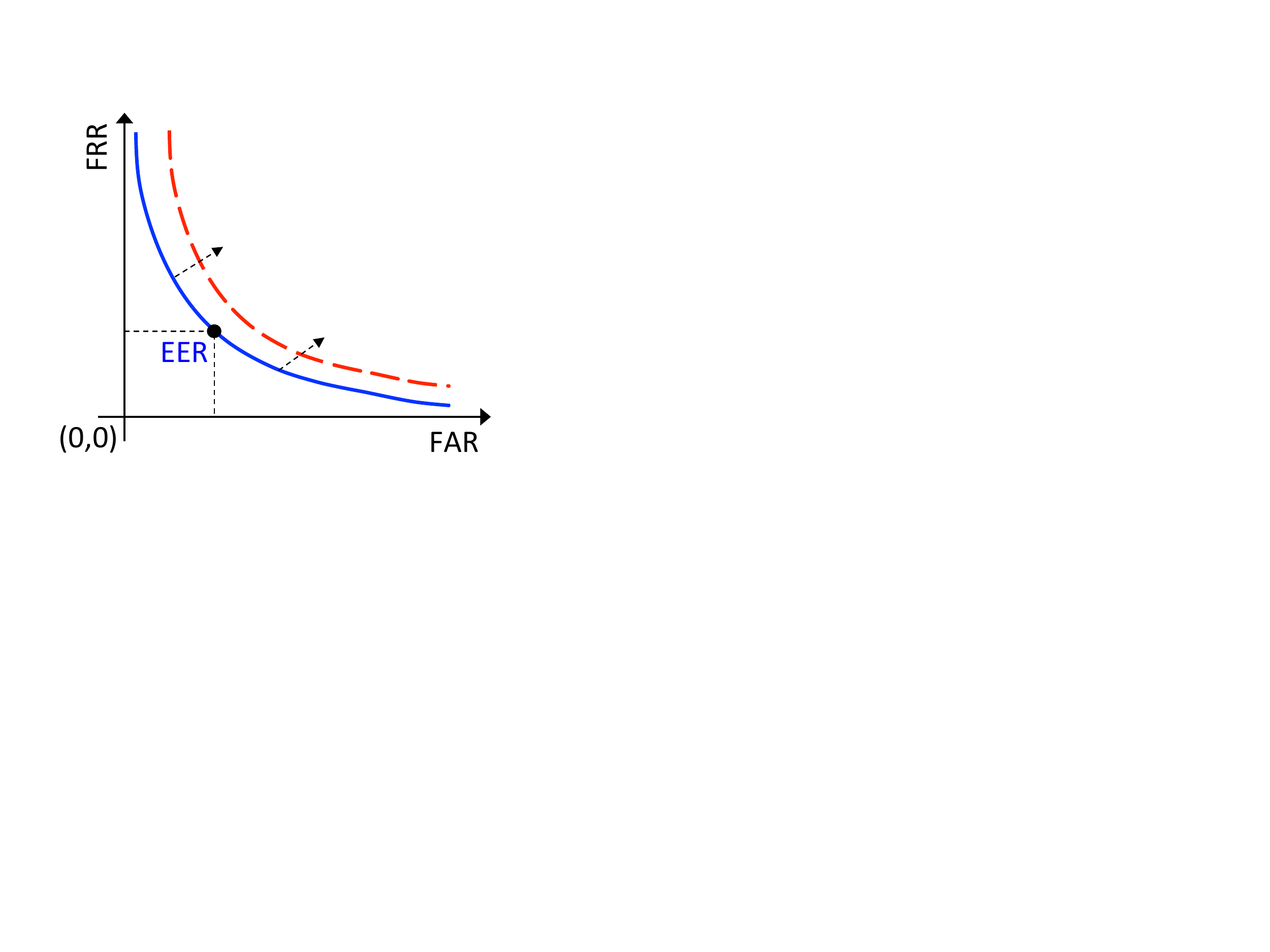}
    \label{fig:ROCcurve}}
\else 
  \subfigure[System Framework]{
    \includegraphics[width=4.9in]{fig1a.pdf}
    \label{fig:generalFramework}}
  \subfigure[ROC Curve]{
    \includegraphics[width=1.9in]{fig1b.pdf}
    \label{fig:ROCcurve}}
\fi
\caption{(a)~Secure biometrics involves encoding the biometric features before storage at the 
access control device. The authentication decision checks whether the probe biometric is 
consistent with the stored data. For clarity, the figure depicts the feature vectors 
extracted from biometrics, rather than the underlying biometric measurements.
(b)~Typical tradeoffs between FAR and FRR in biometric-based authentication systems.  
In general, incorporating security and privacy constraints comes at the price of 
diminished accuracy, which is manifested as a shift of the nominal ROC curve (blue)
away from the axes (red).}
\label{fig:highlevelBD}
\end{figure*}

The model explained above provides a generalized framework within
which to design, evaluate and implement secure biometric
authentication systems. As we shall see later, this framework
accommodates several realizations of secure biometrics.  It can
encapsulate several ways of implementing the encoding and decoding
functions, various biometric modalities, and even different kinds of
adversaries -- computationally unbounded or bounded, possessing side
information or not, and so on. Furthermore, in spite of its
simplicity, the framework permits us to define precisely all
performance measures of interest, including conventional metrics used
to measure accuracy, as well as newer metrics needed to assess
security and privacy.
 
Conventionally two metrics are used to measure the matching accuracy
of biometric systems.  The first is the \emph{False Rejection Rate
(FRR)}, which is the probability with which the system rejects a
genuine user (the missed detection probability).  The second is the
\emph{False Acceptance Rate (FAR)} which is the probability that the
system authenticates a probe biometric that came from a person
different from the enrolled (and claimed) identity. For any given
realization of a biometric access control system, there exists a
tradeoff between these two quantities as illustrated in
Figure~\ref{fig:ROCcurve}.  It is not possible simultaneously to
reduce both beyond a fundamental limit governed by the statistical
variations of biometric signals across users and measurement noise and
uncertainties.  The performance of a biometric access control system
is typically characterized by its empirical Receiver Operating
Characteristic (ROC) curve which is a plot of the empirical FRR
against the empirical FAR. Based on the ROC curve, the performance is
sometimes expressed in terms of a single number called the \emph{Equal
Error Rate (EER)} which is the operating point at which FAR equals
the FRR, as is depicted in Figure~\ref{fig:ROCcurve}.
  
In addition to the two conventional metrics discussed above, in the
following we present three performance measures that allow us to
characterize the privacy, security and storage requirements of a
secure biometric system.

The first is \emph{Privacy Leakage}.  This is the number of bits of
information leaked about the biometric feature vector $\mathbf{A}$
when an adversary compromises the stored data $\mathbf{S}$
and/or the secret key $\mathbf{K}$. An information
theoretic measure of privacy leakage is mutual information
$I(\mathbf{A};\mathbf{V}) = H(\mathbf{A}) - H(\mathbf{A}|\mathbf{V})$,
where $\mathbf{V}$ represents the information
compromised by the adversary and may equal $\mathbf{S}$,
$\mathbf{K}$, or the pair $(\mathbf{S}, \mathbf{K})$.
The two terms on the right hand side are, respectively, the entropy of $\mathbf{A}$ and the conditional entropy (or ``equivocation'') of $\mathbf{A}$ given the leaked data $\mathbf{V}$. As $H(\mathbf{A})$ quantifies the number of bits required to specify $\mathbf{A}$ and $H(\mathbf{A} | \mathbf{V})$ quantifies the remaining uncertainty about $\mathbf{A}$ given knowledge of $\mathbf{V}$, the mutual information is the {\em reduction} in uncertainty about $\mathbf{A}$ given $\mathbf{V}$~\cite{Cover91B}.
Mutual information (or equivalently, equivocation) provides a strong 
characterization of the privacy leakage~\cite{lai11tifs1,lai11tifs2,ignatenko09tifs,wang12tifs}. 

The accuracy with which an adversary can {\em reconstruct} the original biometric is 
often used as an additional performance metric~\cite{talwai11icassp},
and sometimes as a loose proxy for privacy leakage. Driving privacy leakage (as
defined by mutual information) to zero necessarily maximizes the adversary's reconstruction 
distortion. This is due to the data processing inequality and the rate-distortion theorem of 
information theory \cite{Cover91B}. However, for many commonly encountered distortion 
functions that measure the average distortion per component, e.g., the normalized Hamming distortion,
the reverse is not true, i.e., maximizing the adversary's distortion does not necessarily minimize privacy leakage in terms of mutual information. To illustrate how this could happen, consider a scheme which reveals to the adversary that the user's (binary) biometric feature vector is equally likely to be one of two possibilities: the true vector or its bit-wise negation. The adversary's best guess would get all bits correct with probability 0.5 and all incorrect with probability 0.5. Thus, under a normalized Hamming distortion measure, the expected distortion would be 0.5, i.e., the same as guessing each bit at random.  However, while the expected distortion is maximum, all but one bit of information about the biometric is leaked. The mutual information measure would indicate this high rate of privacy leakage. Thus reconstruction distortion cannot be a proxy for privacy leakage although the two are loosely related as discussed above.

The second performance measure is the \emph{Successful Attack Rate (SAR)}.  This is the
probability with which a system authenticates an adversary instead of
the victim, where the adversary's knowledge has been enhanced by some
side information consisting of the victim's
biometric, stored data, and/or key. The SAR is
always greater than or equal to the nominal FAR of the system.
This follows because the side information can only improve the
adversary's ability to falsely authenticate. 

The above definition of security is different from
that used in some of the literature.  Our definition of SAR is
specific to the authentication problem, quantifying the probability
that an adversary gains access to the system.  In other settings,
security has been related to the difficulty faced by an adversary in
discovering a secret that is either a function of the biometric, or
is chosen at enrollment, see,
e.g.,~\cite{lai11tifs1,lai11tifs2,ignatenko09tifs}.  The motivation
for using SAR as the security metric in our development is
two-fold. First, as in~\cite{lai11tifs1,lai11tifs2,ignatenko09tifs},
it can capture the difficulty of discovering the user's secret and
thereby gaining access to the system. Second, it is conceptually
related to the FAR; the SAR defaults to the FAR when the adversary has no
side information.  Given a choice of two systems, and knowledge of
the possible attack scenarios, a system designer may prefer the
system with the higher FAR if it provides the lower SAR of the two.

The third and final measure is the \emph{Storage Requirement} per
biometric.  This is the number of bits needed to store $\mathbf{S}$
and, in two-factor systems, $\mathbf{K}$.  For some secure biometrics
realizations, this can be much smaller than the number of bits used to
represent $\mathbf{A}$. For methods involving encryption, this value
can be much larger owing to ciphertext expansion. For detailed
mathematical definitions of these metrics, we refer the reader to
\cite{wang12tifs}.
 
Unlike the FAR/FRR tradeoff which has been extensively studied, the inter-relationships between privacy 
leakage, SAR and the FAR/FRR performance are less clearly understood. It is important to realize that privacy leakage 
and security compromise (quantified by the SAR) 
characterize distinct adversarial objectives: An adversary may
discover the user's biometric without necessarily being able to break
into the system. Alternatively, an adversary may illegally access the
system without necessarily being able to discover the user's
biometric.

\section{Processing of Biometric Signals}

\begin{figure}[t]
\centering
\includegraphics[width=3in]{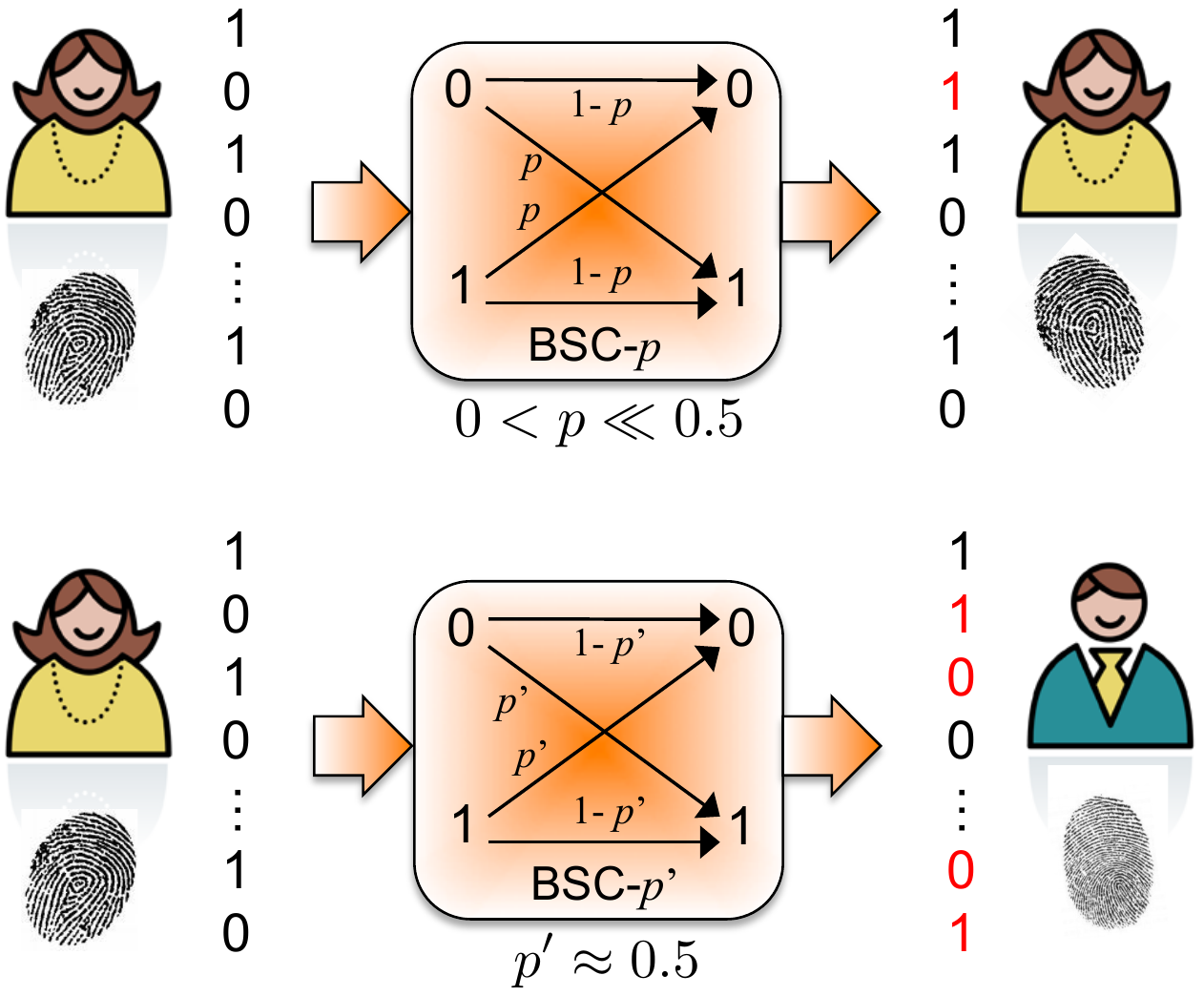}
\caption{Binary feature vectors extracted from two biometric
  measurements can be related by a Binary Symmetric
  Channel (BSC). A good feature extraction algorithm ensures that the
  crossover probability is low when the measurements come from the
  same user and nearly 0.5 if the measurements come from different
  users.}
\label{fig:BSCmodel}
\end{figure}

Let us first consider the properties of biometric feature vectors that
would ensure good accuracy, i.e., a low FRR and a low FAR. It is often
useful to think about biometric variability in terms of
communications: any two biometric measurements can be regarded as the
input and output of a communication channel. If the measurements are
taken from the same user, they will typically be quite similar, and
the channel has little ``noise''. In contrast, if the measurements
come from different users, they will typically be quite different, and
the channel noise will be large. A ``good'' feature
extraction algorithm must deliver this type of variability among
biometric samples -- strong intra-user dependence and weak
inter-user dependence.  A simple case is binary features where the
relationship between feature vectors can be modeled as a binary
bit-flipping (``binary-symmetric'') channel. This
is depicted in Figure~\ref{fig:BSCmodel} where the crossover
probability between feature bits of the same user is small ($0 < p
\ll 0.5$), and that between feature bits of different users is large
($p^{\prime}\approx 0.5$).  Smaller feature vectors are also
desirable due to lower storage overhead.

In practice, the situation is more complicated: 
the statistical variation between biometric measurements is user specific, i.e., 
some users inherently provide more strongly correlated measurements than others~\cite{doddington98icslp}. 
Furthermore, depending upon the feature 
extraction algorithm, some elements of a feature vector may remain more stable 
across multiple measurements than others~\cite{wang09wifs}. The statistical
variation is typically estimated at the enrollment stage by taking several samples from 
the individual being enrolled. This allows the system designer to
set (possibly user-specific) parameters, e.g., acceptance thresholds, to accommodate
the typical variation between 
enrollment and probe biometrics. Biometric feature extraction is a rich area
of research, and several algorithms have been proposed for
extracting discriminable information from
fingerprints~\cite{maltoni03B}, irises~\cite{daugman04csvt,daugman06procieee}, faces~\cite{zhao03csur,bowyer06cviu,li11B}, speech~\cite{reynolds95tsap} and more 
exotic biometric modalities such
as gait~\cite{boulgouris05spm,nixon06procieee} and ECGs~\cite{israel05pr}.

\begin{figure}[t]
\centering
\includegraphics[width=3.5in]{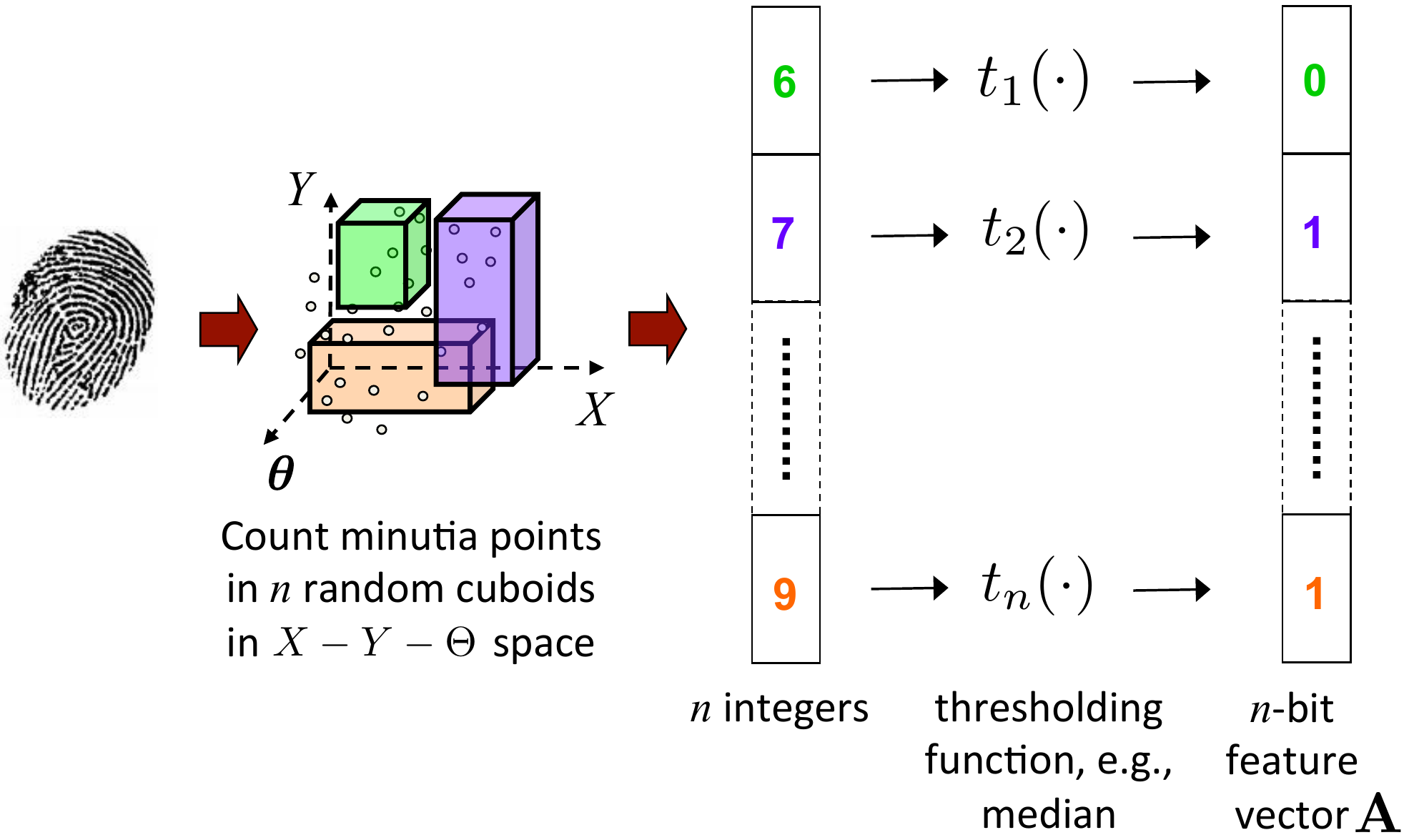}
\caption{Each random cuboid in the $X-Y-\Theta$ space contributes one bit toward an 
$n-$bit binary feature vector. A thresholding function converts the $n$-length integer vector of 
minutia counts to a binary feature vector. An example of a  threshold for each cuboid is the 
median of minutia counts computed over 
all enrollment fingerprints of all users in the database. This ensures that each cuboid produces a
`0' bit for half of the fingerprints in the database, and a `1' bit for the other half. This is desirable
because it makes a feature bit maximally hard to guess given no other side information~\cite{sutcu08isit}.}
\label{fig:fe-example}
\end{figure}

In addition to FRR/FAR considerations we can also ask what statistical
properties should the features have to guarantee low privacy leakage?
In the binary example of Figure~\ref{fig:BSCmodel}, it would be
desirable to have the value of a bit at position $i$ in the feature
vector be statistically independent of the value of a bit at position
$j$. This would ensure that a compromised feature bit does not reveal
any information about hitherto uncompromised feature bits. For the
same reason, the value of a feature bit in Alice's feature vector
should ideally be independent of the value of any bit in Bob's feature
vector~\cite{sutcu08isit}. Designing feature vectors to possess such
privacy-preserving properties forces a compromise between
discriminability, i.e., the independence of the feature values, and
robustness, i.e., the reproducibility of the feature values. This, 
in turn, affects the accuracy (FRR and FAR) of the system,
highlighting the fact that privacy comes at the price of performance.


In secure biometrics, the derived features must satisfy an additional
constraint: The operations performed on the features during secure
access control protocols must be permissible within the architecture
of the encoding and decision modules of
Figure~\ref{fig:generalFramework}. For instance, minutia points are the \emph{de facto}
standard features for highly accurate fingerprint matching,
but they cannot directly be encrypted for use in secure biometric
matching, because the mathematics required to model minutiae movement,
deletion and insertion --- such as factor graphs~\cite{vetro09dscbook}
--- are very difficult to implement in the encrypted domain. In
response to this problem, biometrics researchers have used methods
that extract equal-length feature vectors from
biometrics~\cite{jain99cvpr,xu09tifs,bringer10icb,nagar10icassp}. The
idea behind this is to turn biometric matching into a problem of
computing distance (e.g., Euclidean, Hamming, or Manhattan), an operation
that is feasible within secure biometric architectures.
Figure~\ref{fig:fe-example} shows an example in which a
fingerprint impression is transformed into a binary feature vector
that is suitable for Hamming distance-based matching, and amenable to
many secure biometrics architectures~\cite{sutcu08isit}. It must be noted that imposing
constraints on the feature space makes secure architectures feasible,
but forces the designer to accept a degradation in the FAR-versus-FRR
tradeoff in comparison to that which would have been achieved in an
unconstrained setup. This degradation in the tradeoff is depicted in
the ROC curve in Figure~\ref{fig:ROCcurve}. As an example, by fusing
scores from multiple sophisticated fingerprint matchers, it is possible 
for conventional fingerprint access control systems to achieve an EER 
below 0.5\%~\cite{cappelli06pami}. In contrast, to the best of our knowledge, 
no secure fingerprint biometric scheme has yet been reported with 
an EER below 1\%.

%
%

\section{Secure Biometrics Architectures}

We now turn to methods for converting biometric features into
``secure'' signals that can be stored in the biometric database, to be
used for authentication. We briefly cover the four most prominent
classes mentioned in the introduction, treating each as a specific
manifestation of the unified framework of Figure~\ref{fig:generalFramework}.

\subsection{Secure Sketches} \label{sec.secureSketch}

\begin{figure}[t]
\centering
\includegraphics[width=3.3in]{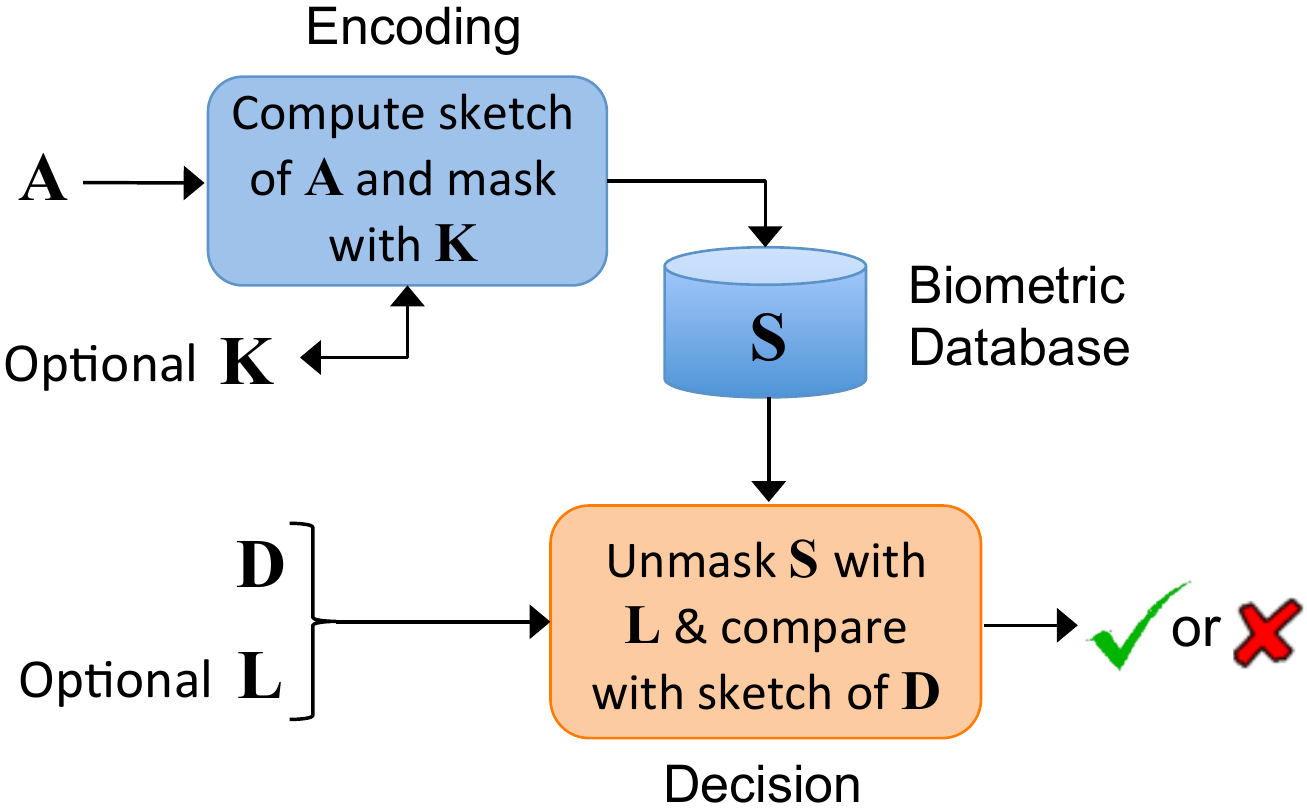}
\caption{In secure sketch systems, encoding involves deriving a ``sketch'' that reveals
little or no information about the underlying biometric. The decision function involves 
determining whether the probe feature vector is consistent with the sketch derived 
from the enrollment feature vector. A two-factor implementation using a secret key
in addition to the biometric features is also possible.}
\label{fig:highlevelSS}
\end{figure}

A secure sketch-based system derives information -- called a
sketch or helper data $\mathbf{S}$ -- from Alice's enrollment
biometric $\mathbf{A}$ and stores it in the access control
database~\cite{dodis04eurocrypt}, as shown in
Figure~\ref{fig:highlevelSS}. The decision function tests whether the
probe biometric $\mathbf{D}$ is consistent with the sketch and grants
access when it is. The sketch $\mathbf{S}$ should be constructed so
that it reveals little or no information about $\mathbf{A}$

Secure sketches can be generated in several ways, for example, by computing
a small number of quantized random projections of a biometric feature
vector~\cite{sutcu07tifs}. A particularly instructive method -- one that shows
the connections between secure sketches and the fuzzy commitment
architecture -- employs error correcting codes (ECCs). The secure
sketch is constructed as a syndrome of an ECC with parity check matrix
$\mathbf{H}$, given by $\mathbf{S} = \mathbf{HA}$.  The idea is that a
legitimate probe biometric $\mathbf{D}=\mathbf{B}$ would be a slightly error
prone version of $\mathbf{A}$. Therefore, authentication can be
accomplished by attempting to decode $\mathbf{A}$ given $\mathbf{D}$
and $\mathbf{S}$.  Secure sketches constructed in this way provide
information theoretic security and privacy guarantees that are
functions of the dimension of the ECC. They also suggest an
interesting interpretation in which $\mathbf{S}$ is a Slepian-Wolf
encoded version of $\mathbf{A}$~\cite{slepian73IT}. Thus, biometric authentication is
akin to Slepian-Wolf decoding~\cite{sutcu08isit}.  This observation
paves the way for implementations based on graphical models, e.g.,
belief propagation decoding coupled with appropriately augmented LDPC
code graphs~\cite{draper07icassp}.

A natural question to ask here is: ``How does the decision box know
that it has correctly decoded $\mathbf{A}$ given $\mathbf{D}$ and
$\mathbf{S}$.'' In practice, this question is answered by storing a
cryptographic hash of $\mathbf{A}$ on the device along with the stored
data $\mathbf{S}$. Then, assuming no hash collisions, if the
cryptographic hash of the decoded vector matches the stored hash, the
device determines that authentication is successful. However, due to
the use of a cryptographic hash, this system is only computationally
secure and not information-theoretically secure.  But, as we will
describe, an information-theoretic test for the recovery of
$\mathbf{A}$ can be constructed by considering the geometry of the
ECC.  The result is an information-theoretically secure solution that
retains the FAR/FRR performance of the design that uses cryptographic
hashes.  This leads to an interesting coding theory exercise, the
details of which are worked out in the sidebar: ``A Linear ECC-based
Secure Sketch Authentication System".
For an implementation of an ECC-based secure sketch-based system
see~\cite{wang09wifs}.  There, using an irregular LDPC code of
length 150 bits, an EER of close to 3\% is achieved.

\addtocounter{bx}{1}
\begin{center}
{ {\bf Begin Sidebar Inset \#\protect\Alph{bx}: A linear
    ECC-based secure sketch authentication system}}
\end{center}
\label{box:illexcsi}

\begin{figure}[t]
\centering
\includegraphics[width=3.3in]{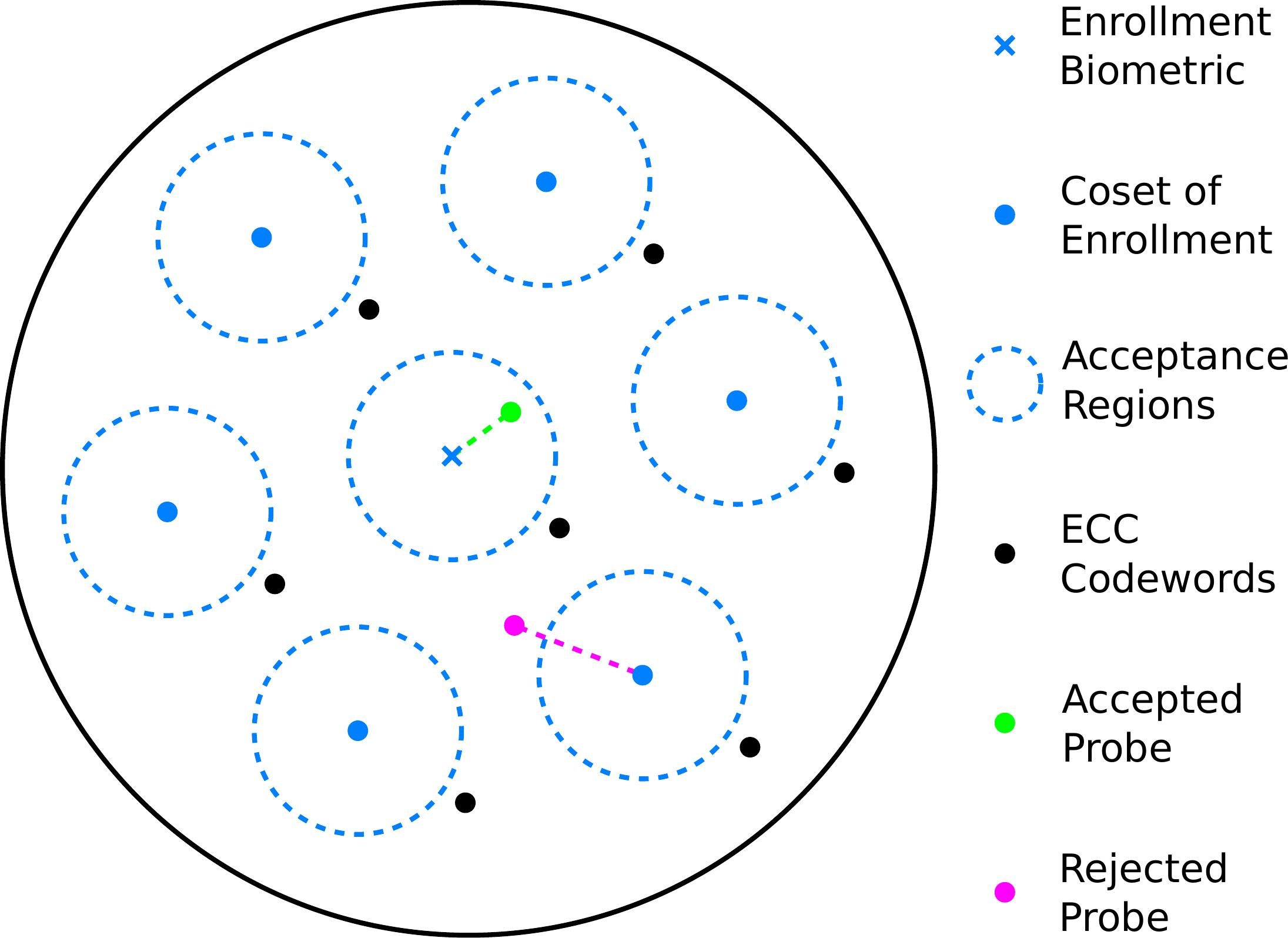}
\caption{An abstract representation of the ECC-based secure sketch authentication system, depicting the acceptance regions in relation to the enrollment biometric and its coset.}
\label{fig:SyndromeDecoding}
\end{figure}
%
%
The underlying geometry of a secure sketch system based on a binary
linear ECC is illustrated in Figure~\ref{fig:SyndromeDecoding}. For
simplicity, we focus on keyless systems, i.e., ones that do not
involve the secret key $\mathbf{K}$. We consider binary biometric
sequences $\mathbf{A}$ of length $n$. The black circle bounds the set
of all length-$n$ binary words. The black dots represent $2^{n-m}$
codewords that correspond to the null space of an $m\times n$ binary
parity check matrix $\mathbf{H}$ of rank $m$.
\\
\\
%
%
\noindent {\textbf{Enrollment via ECC:}} The blue cross 
(arbitrarily placed at the center) indicates the enrollment biometric
feature vector $\mathbf{A}$. Enrollment consists of mapping
$\mathbf{A}$ to its $m$-bit syndrome $\mathbf{S} :=
\mathbf{H}\mathbf{A}$ where operations are in the binary field $\mathbb{F}_2$. 
The set of $2^{n-m}$ binary words that are mapped by $\mathbf{H}$ to $\mathbf{S}$ form the enrollment coset, of which $\mathbf{A}$ is a member.
The blue dots represent the other members of the coset that, together with the blue cross, form the entire enrollment coset.
Knowledge of stored information $S$ is equivalent to knowledge of the members of this coset and hence is available to the decision module.
\\
\\
%
%
\noindent {\textbf{Authentication:}} The first step in authenticating a
probe vector $\mathbf{D}$, is to perform syndrome decoding to recover
$\widehat{\mathbf{A}}$, the estimate of the enrollment vector.  This
is the element of the coset specified by $\mathbf{S}$ that is closest
to $\mathbf{D}$ in Hamming distance. The probe is accepted as
authentic if the normalized Hamming distance between this estimate and
the probe is less than a threshold $\tau$, i.e., $\frac{1}{n}
d_H(\widehat{\mathbf{A}}, \mathbf{D}) < \tau$, where $\tau \in
(p,0.5)$; otherwise the probe is rejected. Thus, the system accepts a
probe if it is within a certain distance of the coset of $\mathbf{S}$,
and otherwise rejects it. In Figure~\ref{fig:SyndromeDecoding}, each
blue dashed circle represents the boundary of an acceptance region
associated with a single coset member that is produced by syndrome
decoding and the threshold test. The overall acceptance region is the
union of these individual acceptance regions.  The green dot is an
example of a probe vector $\mathbf{D}$ that will be accepted and the
magenta dot an example of a $\mathbf{D}$ that will be rejected.
\\
\\
%
%
\noindent {\textbf{FRR:}} The probe $\mathbf{D} = \mathbf{B}$ of a
legitimate user is a noisy version of $\mathbf{A}$.  Ideally this is
equivalent to the output of a BSC-$p$ channel with input $\mathbf{A}$,
so that the bits of the noise vector $(\mathbf{A} \oplus \mathbf{B})$
 are independent Bernoulli-$p$ random variables independent of
$\mathbf{A}$. For any $\mathbf{A}$, the FRR is equal to the
probability that the noise pushes it outside the acceptance
region. Since the code is linear, all cosets are translations of the 
coset of all codewords whose syndrome is zero.  Hence the FRR is the
same for all $\mathbf{A}$. It turns out that $\mathbf{H}$ can be
designed to make FRR exponentially small in $n$ (for large enough $n$)
if (i) the threshold $\tau \in (p, 0.5)$ and (ii) the rate $(n-m)/n$
of the ECC is strictly smaller than the capacity of a BSC-$\tau$
channel~\cite{wang12tifs}.
\\
\\
%
%
\noindent {\textbf{FAR:}} An attacker unassisted by any compromised
information must pick an attack probe uniformly over the space of all
length-$n$ binary words.  The FAR is thus given by the ratio of the total
volume of the acceptance spheres to the overall volume of the
space. Coding theory tells us that, in high dimensions ($n \gg 1$), if 
the rate $(n-m)/n$ of the ECC is strictly smaller than the capacity of a
BSC-$\tau$ channel, the coset members can be
well-separated and the volume outside of the acceptance spheres can be
made to dominate the volume inside them. Thus,
the FAR can also be made exponentially small in $n$ by suitably
designing the ECC~\cite{wang12tifs}.
\\
\\
%
%
\noindent {\textbf{Privacy leakage:}} Knowledge of $\mathbf{S}$ would
reveal to an attacker that $\mathbf{A}$ belongs to the set of
$2^{n-m}$ blue points as opposed to the set of all $2^n$ binary
words. If $\mathbf{A}$ is equally likely to be any $n$-bit word, then
this corresponds to an information-theoretic privacy leakage rate (in
bits) of $I(\mathbf{A}; \mathbf{S}) = m
= \log_2(\#\mbox{all binary sequences}) - \log_2(\#\mbox{blue
points})$.
\\
\\
\noindent {\textbf{SAR:}} From the foregoing discussion, it is clear that an attacker who is
assisted by compromised information (either $\mathbf{A}$ or
$\mathbf{S}$) can determine the acceptance regions and choose an
attack probe that falls within them.  Thus, given such side
information, the SAR of this system is one. This property, however, is
not unique to this system, but a general drawback of {\it any keyless
system} \cite{wang12tifs}.  A two-factor scheme partially addresses
this drawback by using a key $\mathbf{K}$ independent of $\mathbf{A}$
in the enrollment state, keeping SAR down to the nominal FAR when
$\mathbf{A}$ is compromised. However, revealing $\mathbf{S}$ to the
adversary still results in SAR = 1.

\begin{center}
{ {\bf End Sidebar Inset \#\protect\Alph{bx} }}
\end{center}

The preceding explanation assumes a keyless secure sketch architecture.
However, as shown in Figure~\ref{fig:highlevelSS}, a two-factor implementation is
possible by using an independent key $\mathbf{K}$ provided by the system or
chosen by the user. The advantage of the two-factor architecture is enhanced privacy 
and security, as well as revocability of a compromised secure biometric $\mathbf{S}$ or key 
$\mathbf{K}$. Specifically, when the adversary discovers either $\mathbf{K}$ or $\mathbf{S}$, but not both, 
the two-factor system suffers no privacy leakage. Furthermore, when the adversary discovers $\mathbf{K}$ or the
biometric $\mathbf{A}$, but not both, the SAR is still no larger than the nominal FAR of the 
system~\cite{wang12tifs}. In other words, the second factor $\mathbf{K}$ prevents privacy
leakage while preventing degradation in the biometric authentication performance~\cite{ignatenko09tifs}.
Lastly, if only either $\mathbf{K}$ or $\mathbf{S}$ is compromised by 
an attacker, the enrollment can be revoked by discarding the other factor. The user can then 
refresh their enrollment without any security or privacy degradation.
The penalty of two-factor system is a loss of convenience, since the user must either memorize $\mathbf{K}$ or carry it on a smart card.

\subsection{Fuzzy Commitment}

\begin{figure}[t]
\centering
\includegraphics[width=3.3in]{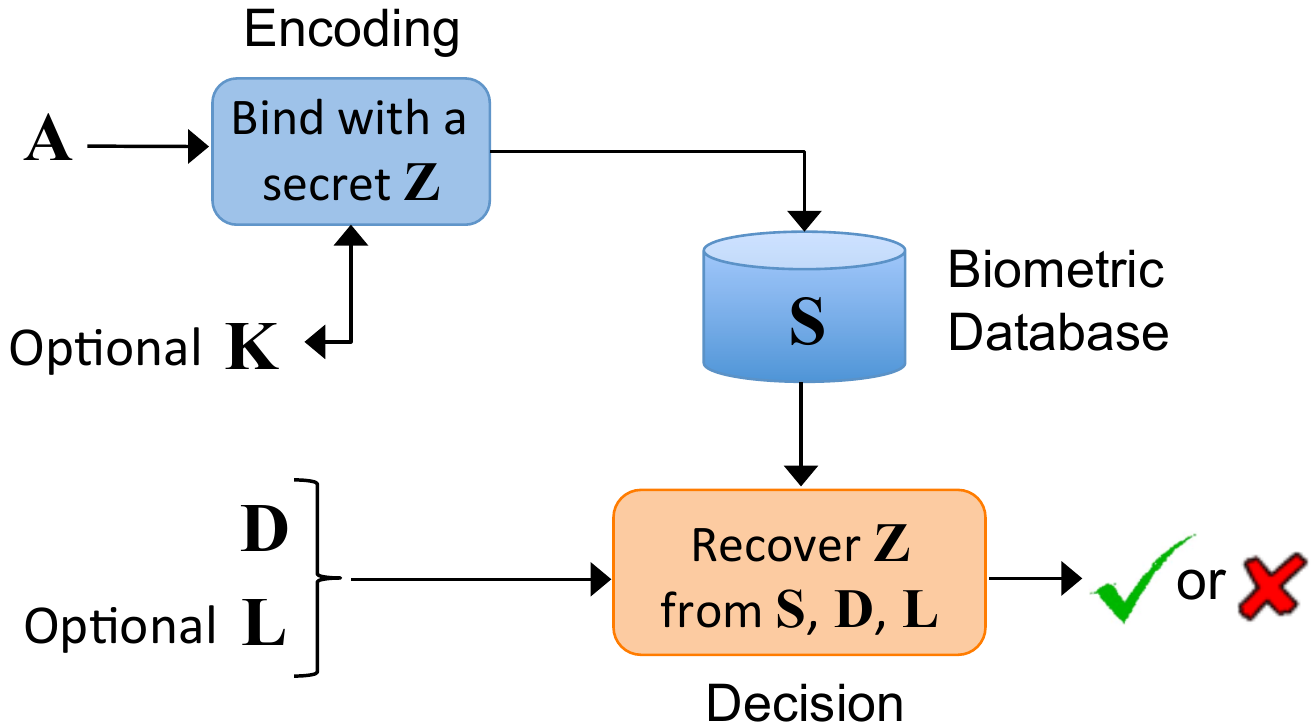}
\caption{In fuzzy commitment, encoding involves binding the biometric features to a randomly
generated vector $\mathbf{Z}$ resulting in stored data $\mathbf{S}$. The decision module checks 
whether $\mathbf{Z}$ is exactly recovered using the probe feature vector and the stored data.
A two-factor realization with a user-specific key in addition to the biometric feature is also 
possible.}
\label{fig:highlevelFC}
\end{figure}

Fuzzy commitment involves binding a secret message to the enrollment biometric which can later be recovered 
with a legitimate probe biometric to perform authentication~\cite{juels99fuzzy,juels06jdcc}.
As depicted in Figure~\ref{fig:highlevelFC}, Alice binds her biometric feature vector $\mathbf{A}$ to a randomly
generated vector $\mathbf{Z}$, producing the data $\mathbf{S}$ which is stored in a database as the secure 
biometric. Again, the encoding function should ensure that $\mathbf{S}$ leaks little or no information about
$\mathbf{A}$ or $\mathbf{Z}$. To perform authentication, a user claiming to be Alice provides a 
probe biometric feature vector $\mathbf{D}$ and the device attempts to recover $\mathbf{Z}$. Access is 
granted only when there is exact recovery of the message $\mathbf{Z}$, which would happen only if $\mathbf{D}$
is sufficiently similar to $\mathbf{A}$.

There are several ways to bind a secret message to the enrollment biometric. One such
method uses quantization index modulation (QIM)~\cite{chen01IT}, in which the biometric
features are quantized in such a way that the choice of the quantizer is 
driven by the secret message~\cite{bui2010tifs}. Another method uses 
error correcting codes. We explain this ECC embodiment below
because it clarifies the basic concepts of fuzzy commitment using familiar ideas from
channel coding. Assuming that all vectors are binary, consider a simple
example wherein the secure biometric is computed as $\mathbf{S} =
\mathbf{G}^{T} \mathbf{Z} \oplus \mathbf{A}$, where $\mathbf{G}$ is
the generator matrix of an ECC. During authentication, the access
control device receives the probe vector $\mathbf{D}$ and computes
$\mathbf{S} \oplus \mathbf{D}$ which results in a noisy codeword. The
noise is contributed by the difference between $\mathbf{A}$ and
$\mathbf{D}$.  Then, using classical ECC decoding, the device attempts
to decode the random message $\mathbf{Z}$ and allows access only if it
is successful. The ECC-based implementation provides concrete
information theoretic guarantees of privacy and security depending
upon the parameters of the selected ECC. In fact, in terms of the FRR, FAR, 
privacy leakage, and SAR this
ECC-based construction is equivalent to the ECC-based secure sketch
construction discussed earlier~\cite{wang12tifs}. They are not
identical however, as the storage requirement of fuzzy commitment is generally 
greater than that of secure sketch.

An alternative way of understanding the ECC-based implementation of fuzzy commitment is
to view it as a method of extracting a secret $\mathbf{Z}$ by means of polynomial 
interpolation~\cite{juels99fuzzy,juels06jdcc}. Suppose that the decoder is given a large constellation of candidate
feature points (vectors) containing a few genuine points and a large number of ``chaff'' points, generated
for the purpose of hiding the relevant points. The secret can be recovered only by interpolating
a specific polynomial that passes through the relevant feature points for the user being tested. It is inefficient to 
perform polynomial interpolation by brute force. Fortunately, polynomial interpolation can be efficiently accomplished
by ECC decoding, for example, Reed-Solomon decoding using the Berlekamp-Massey algorithm~\cite{massey69it}.
This realization has inspired many implementations of fuzzy commitment, primarily for fingerprints,
where polynomial interpolation is applied to a collection of genuine and chaff points constructed from
locations and orientations of fingerprint minutiae~\cite{clancy03acmbma,yang05icassp,nandakumar07tifs,nagar08icpr}. An example implementation
of such a fuzzy commitment scheme appears in~\cite{nagar08icpr}, wherein a (511,19) BCH code
is employed for polynomial interpolation; experiments show that when the degree of the interpolated polynomial
is increased, the matching becomes more stringent, reducing the FAR, but increasing the FRR.

Based on the relationships between the ECC-based constructions discussed so far, it 
becomes clear that the fuzzy commitment is closely related to the secure sketches.
In fact, it is possible to show that if a secure sketch scheme is given, it can be used to 
construct a fuzzy commitment scheme~\cite{dodis04eurocrypt}. As explained in the case of secure sketch, 
in practical systems, a cryptographic hash of $\mathbf{Z}$ is stored on the device 
along with $\mathbf{S}$, to verify correct recovery 
of $\mathbf{Z}$. Furthermore, a two-factor scheme that utilizes a key $\mathbf{K}$ independent of the enrollment vector $\mathbf{A}$ can similarly improve security and privacy performance, as well as enable revocability.

\subsection{Biometrics as Secure Multiparty Computation}

\begin{figure}[t]
\centering
\includegraphics[width=3.2in]{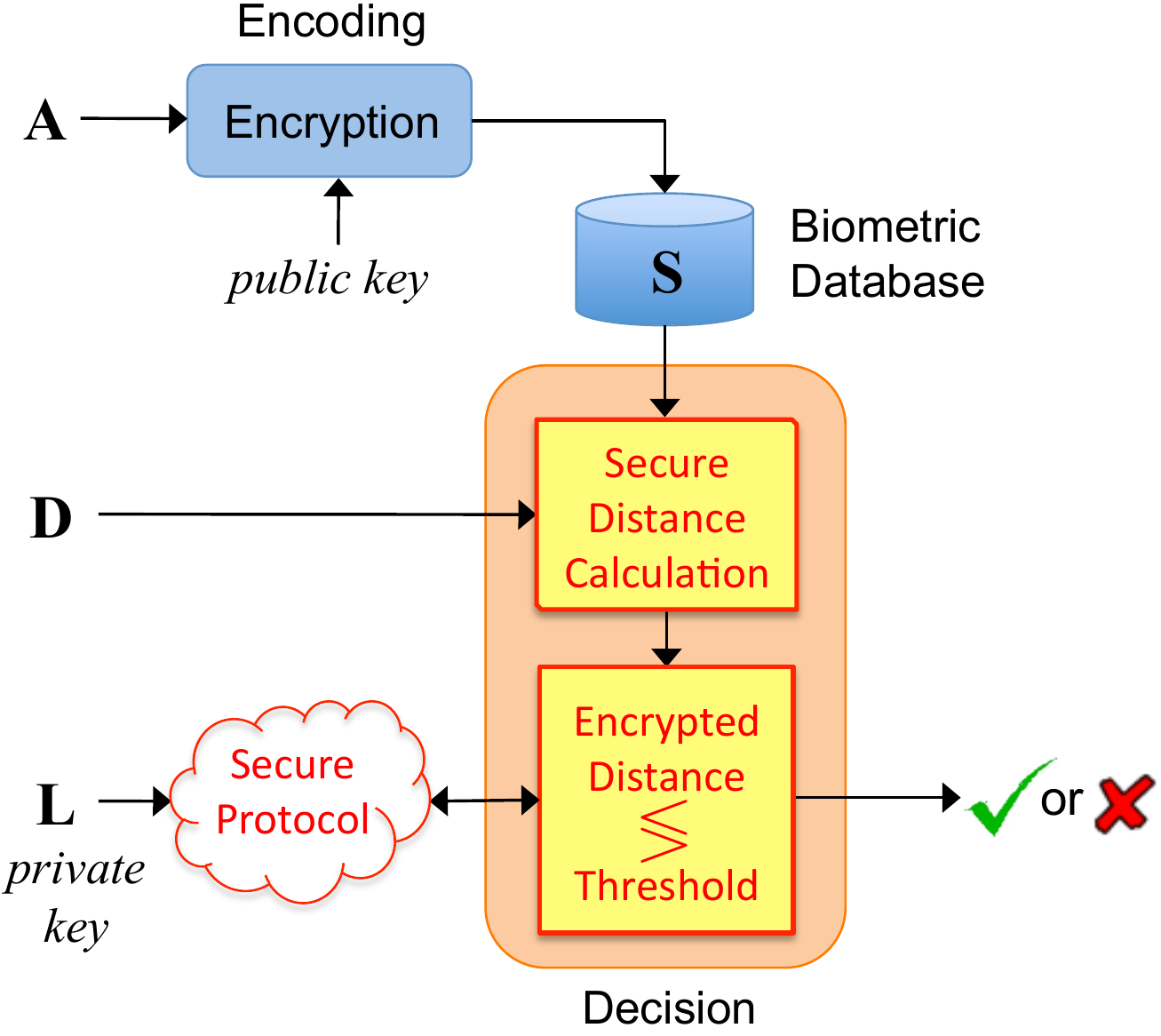}
\caption{In biometrics based on multiparty computation, enrollment involves encrypting the 
biometric features. The authentication decision involves encrypted-domain distance computation
and followed by a comparison protocol between the claimant, who possesses a secret
decryption key $\mathbf{L}$ and the database server which only sees encrypted data.}
\label{fig:highlevelSMC}
\end{figure}

This architecture involves finding the distance between enrollment
and probe biometric features in the encrypted domain. There has been
intense research activity recently on accomplishing this using
public-key homomorphic cryptosystems. These allow an operation on the
underlying plaintexts --- such as addition or multiplication --- to be
carried out by performing a suitable operation on the ciphertexts. To
fix ideas, consider the following simple example.  Suppose the
length-$n$ enrollment feature vector $\mathbf{A}$ is encrypted
elementwise using an additively homomorphic cryptosystem and the
resulting ciphertext $\mathbf{S}$ is stored in the database of the
access control system, as shown in Figure~\ref{fig:highlevelSMC}. An
additively homomorphic cryptosystem, e.g., the Paillier cryptosystem~\cite{paillier99eurocrypt},
satisfies $E(a)E(b) = E(a + b)$ for integers $a,b$ and encryption
function $E(\cdot)$.

A realistic assumption in our simple example is that the encryption key is public, while the 
decryption key $\mathbf{L}$ is available only to the individual attempting to authenticate. Thus, by construction,
this secure biometrics architecture results in two-factor systems, in which the first factor is a 
biometric token and the second factor is a privately held decryption key for a homomorphic
cryptosystem. Suppose a user claiming to be Alice (say) provides a probe 
feature vector $\mathbf{D}$ for authentication. Since the encryption key is public, the device 
can encrypt elements of probe biometric $\mathbf{D}$ and compute the squared distance 
between $\mathbf{A}$ and $\mathbf{D}$ in the encrypted domain using the additively 
homomorphic property as:
\ifonecol 
\begin{align}
&E\left(\sum_{i=1}^n a_i^2 \right) E \left(\sum_{i=1}^n d_i^2 \right) \prod_{i=1}^n E(a_i)^{-2 d_i}  
= E \left( \sum_{i=1}^n (a_i - d_i)^2 \right) \notag.
\end{align}
\else 
\begin{align}
&E\left(\sum_{i=1}^n a_i^2 \right) E \left(\sum_{i=1}^n d_i^2 \right) \prod_{i=1}^n E(a_i)^{-2 d_i}  \notag \\
&= E \left( \sum_{i=1}^n (a_i - d_i)^2 \right) \notag.
\end{align}
\fi
The device then executes a privacy-preserving comparison protocol with
the user to be authenticated to determine whether the distance is
below a threshold. The protocol ensures that the claimant does not
discover the threshold, while neither the claimant nor the device
discovers the actual value of the distance or any of the $a_i$. If the
distance is below the threshold, access is granted. Clearly, the
claimant -- whether Alice or an adversary --- must use the correct
decryption key, otherwise the protocol will generate garbage values.

The example above is meant to illustrate the basic concepts of secure
biometrics based on multiparty computation. Many extensions of the
above scheme have been studied, all of which involve some form of
privacy-preserving nearest neighbor computation~\cite{barni10mmsec,blanton11esorics,bringer07isp,rane13spm}. The
protocols apply a combination of homomorphic encryption and garbled
circuits.  The latter is especially useful in the final
authentication step, i.e., performing an encrypted-domain comparison
of the distance between the enrollment and probe biometrics against a
predetermined threshold. The distance measures need not be restricted
to Euclidean distance; secure biometric comparisons based on 
Hamming distance and Manhattan ($\ell_1$) distance have also been
realized. Privacy-preserving nearest-neighbor protocols such as these
have been proposed for various biometric modalities, for instance,
face images~\cite{sadeghi09icisc}, fingerprints~\cite{barni10mmsec}
and irises~\cite{blanton11esorics}. For further details and analysis of the steps
involved in the cryptographic protocols for biometric authentication,
we refer the reader to a recently published survey
article~\cite{lagendijk13spm}.

Privacy and security in these methods depend on proper protocol design
and key management to ensure that the attacker does not gain access to
the decryption keys.  Privacy and security guarantees are
computational, not information-theoretic, i.e., they rely on the
unproven hardness of problems such as factorization of large numbers,
the quadratic residuosity problem~\cite{paillier99eurocrypt}, or
the discrete logarithm problem~\cite{menezes96book}. In other words,
if Alice's decryption key is discovered by an adversary, then the
system becomes vulnerable to a wide variety of
attacks. Depending upon his computational resources, the adversary can
now query the system using several (possibly synthetic) candidate
biometrics until access is granted by the system, thereby resulting in
a successful attack. Further, the adversary gains a
reasonable proxy biometric vector to be used in the future to
impersonate Alice. Even though it is difficult to give concrete
expressions for the SAR and the privacy leakage for such a system, it
is clear that a compromised decryption key will significantly
increase both the SAR and the privacy leakage.

This architecture requires the database server to store encryptions
of biometric features, therefore the storage cost is high owing to
ciphertext expansion. This is because of the large key sizes used in the
privacy-preserving protocols; typical values are 1024 or 2048 
bits~\cite{lagendijk13spm}. The computational complexity is also much higher
than the other architectures due to the high overhead of interactive
encrypted-domain protocols. 

\subsection{Cancelable Biometrics}

\begin{figure}[t]
\centering
\includegraphics[width=3.2in]{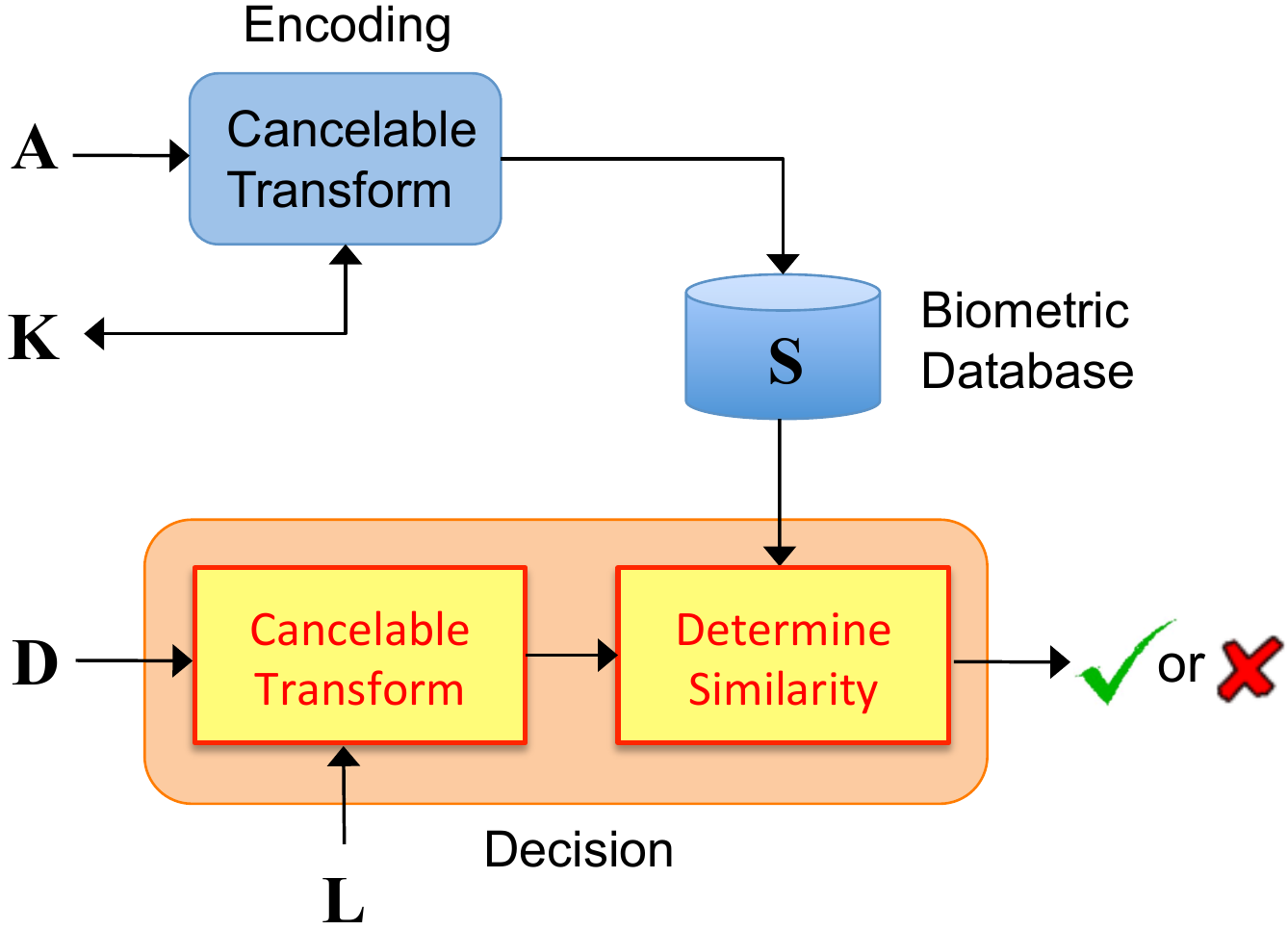}
\caption{In cancelable biometrics, encoding involves applying a secret distorting transform indexed
by a key $\mathbf{K}$ to generate the stored data $\mathbf{S}$. The decision function involves applying a distorting transform, indexed by a key $\mathbf{L}$, to the probe biometric $\mathbf{D}$
and determining whether the result is sufficiently similar to $\mathbf{S}$.}
\label{fig:highlevelCB}
\end{figure}

\begin{figure}[t]
\centering
\includegraphics[width=3.2in]{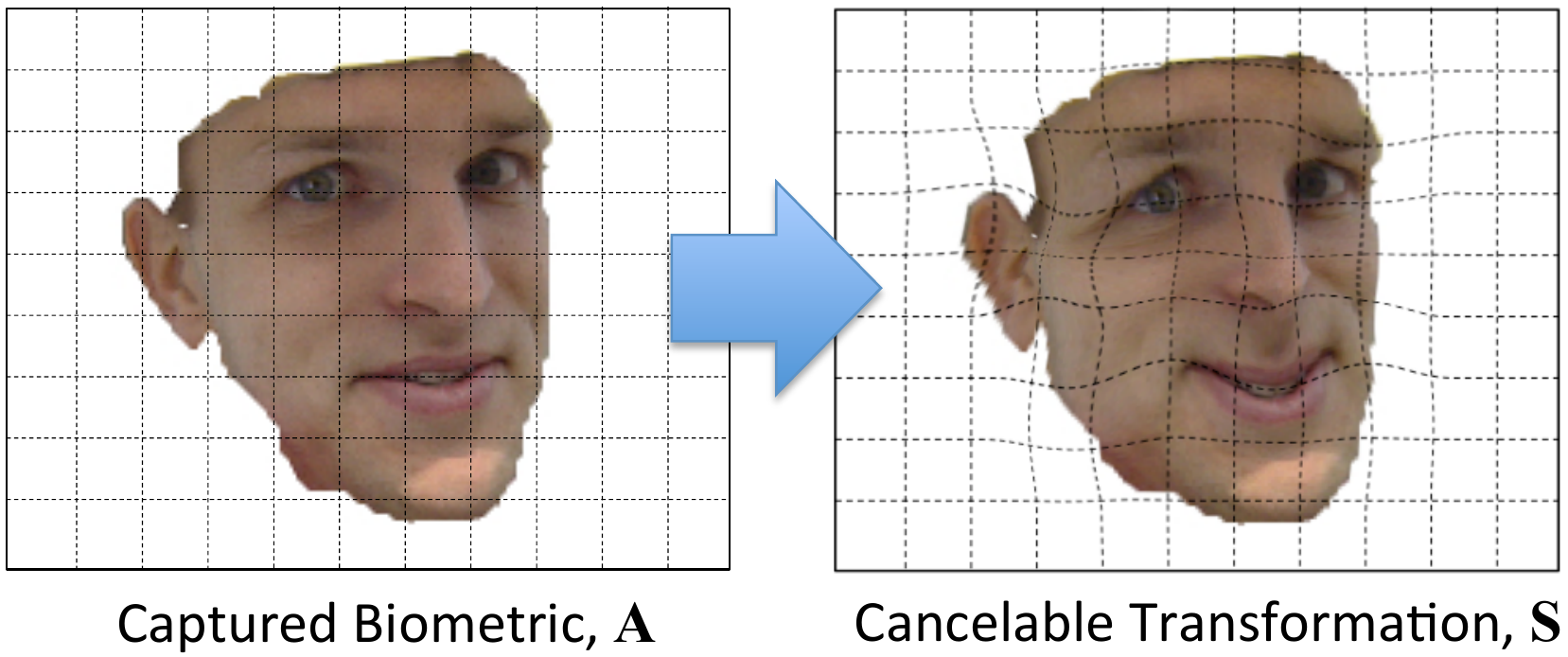}
\caption{An example of a cancelable transformation of a face image. If the stored
data $\mathbf{S}$ is known to have been compromised, the system administrator can
revoke it, and store a different transformation as the new enrollment.}
\label{fig:foremanface}
\end{figure}

Cancelable biometrics refers to a class of techniques in which the enrollment biometric signal 
is intentionally distorted before it is stored in the biometric 
database~\cite{ratha01ibm}. This architecture is depicted in Figure~\ref{fig:highlevelCB}.
The distorting function is repeatable, so that it can be applied again to the probe biometric, 
facilitating comparison with the distorted enrollment biometric. Further, the distorting function is intended to
be a non-invertible and ``revocable'' mapping. This means that, if Alice's stored distorted 
biometric is known to have been compromised, a 
system administrator can cancel her enrollment data, apply a fresh distorting function to 
Alice's biometric, and store the result as her new enrollment. 

The most popular methods of implementing cancelable biometrics involve 
non-invertible mappings applied to rectangular tessellations of face or fingerprint images
\cite{ratha01ibm}, salting of biometric features with a secret 
key~\cite{kong06}, and computing quantized random projections of biometric feature
vectors~\cite{teoh06infoprocletters}.  An example of a cancelable transformation
applied to a face image, similar to schemes proposed in~\cite{ratha01ibm}, is shown in 
Figure~\ref{fig:foremanface}. To authenticate in this
architecture, a user must provide their biometric measurement along with correct 
distorting transformation that should be applied to the measurement. Thus, by construction, 
these are two-factor systems in which the second factor $\mathbf{K}$
is a secret value held by the user which indexes the user-specific deformation, or
salting key, or the realization of a random matrix. The secret value can be in the form
of a memorized PIN number or a longer key held on a smart card.

As would be expected, the choice of the space of 
distorting functions affects the accuracy, i.e., the FAR, FRR and EER for the system under
consideration. The non-invertibility of the distorting function ensures that an adversary cannot
recover the underlying biometric by reading the database of distorted biometrics. In other
words, privacy leakage can be low, or zero, depending on the implementation. Most importantly, 
the secrecy of the 
chosen distorting function is critical as far as the SAR is concerned.
In the various cancelable biometrics
implementations, if the chosen non-invertible image transform,  or the salting key, or the 
realization of the random projection matrix  are revealed, the adversary's task is considerably 
simplified: he needs to find some biometric signal that, when transformed according to the
revealed distorting function, yields an output that is similar to the stored enrollment
data. This would be sufficient for the adversary to gain unauthorized access. 

Though many cancelable transformations have been proposed, formal
proofs regarding the accuracy, privacy and security of these methods
are elusive.  In other words, given a distorted biometric database, we
do not always have a quantitative measure of how difficult it is to 
discover the distorting function and subsequently compromise a legitimate
user's biometric.  Further, even given the distorting function, indexed by the secret key $\mathbf{K}$,
we do not always have a quantitative measure of how easy it is to gain unauthorized
access to the system. Finally, it is not always possible to quantify
the degradation (if any) in the FAR-versus-FRR performance when when a
user's enrollment data is repeatedly revoked and
reassigned. Nevertheless, the low implementation complexity, the large
variety of distorting transformations, and the conceptual simplicity
of the administrative tasks needed to revoke compromised templates
makes cancelable biometrics an attractive architecture.  This is
especially true in scenarios in which a user has enrolled the same
biometric -- e.g., her index finger -- at multiple access control
devices.

 \section{Multiple Secure Biometric Systems}

\begin{figure}[t]
\centering{
\includegraphics[width=3.3in]{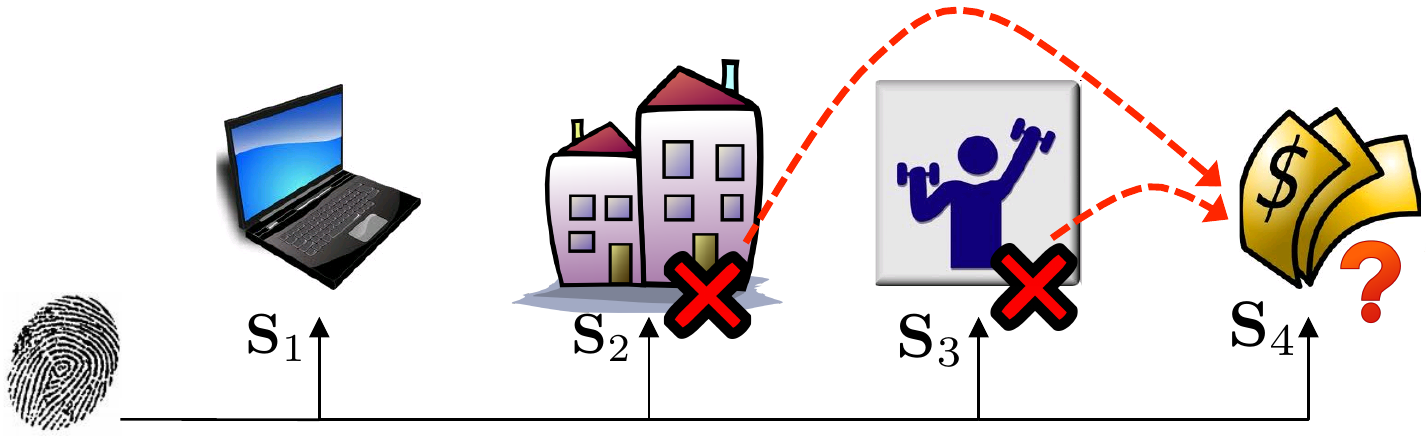}
\caption{An adversary can compromise security and privacy by attacking
multiple devices at which the victim is enrolled.}
\label{fig:multiple-secbio-cartoon}}
\end{figure}

We now consider a topic that is extremely important but remains little
investigated, namely the implications for security and privacy when a
user has enrolled a biometric on several access control
devices. As a concrete example, say that Alice has enrolled her
fingerprints at her bank, at her gym, on her laptop, and at her
apartment complex. In this case, an adversary may first attempt to
compromise the systems that have less stringent security requirements,
perhaps the apartment complex and/or the gym, as shown in 
Figure~\ref{fig:multiple-secbio-cartoon}.  The adversary could
then use the information acquired to attack the more sensitive
systems; for instance to gain access to Alice's bank accounts.
There is an inherent tension between the need for security from an attack spanning multiple systems and the desire to preserve as much privacy as possible in the face of one or more systems being compromised.
We illustrate this tradeoff with a simplified example in the sidebar on
``Tradeoff between security and privacy leakage in multiple biometric systems''.

\begin{center}
{\bf Begin Sidebar B: Tradeoff 
  between security and privacy leakage in multiple biometric systems}
\end{center}

We use the secure sketch architecture for this discussion.  Let the binary ECC
used in enrollment be of length four and span a subspace of
dimension two.  This means there are $2^2 = 4$ codewords.  We consider
the $[4,2]$ code described by the parity-check matrix $\mathbf{H}$ where
\begin{equation*}
\mathbf{H} = \left[\begin{array}{cccc} 1 & 0 & 1 & 1 \\ 0 & 1 & 1 & 1
    \end{array} \right].
\end{equation*}
The codewords of this code are
$\mathcal{C} = \{[0000]^T, [1110]^T, [1101]^T, [0011]^T\}$.
As an example of an enrollment let $\mathbf{A} =
[1 0 1 1]^T$, yielding as stored data the syndrome $\mathbf{S} = \mathbf{H} \mathbf{A} =
[1 \, 0]^T$.  The set of candidate biometrics that share this syndrome
(coset members) are $\mathcal{P} = \{[1011]^T, [0101]^T, [0110]^T,[1000]^T\}$.
For simplicity, we set the decision threshold for
authentication, i.e., the radius of the blue dashed circles in
Figure~\ref{fig:SyndromeDecoding}, to be $\tau = 0$. In this case, access 
will be given only if the probe $\mathbf{D} \in \mathcal{P}$.
  
\begin{figure}[t]
\centering{
\includegraphics[width=3.5in]{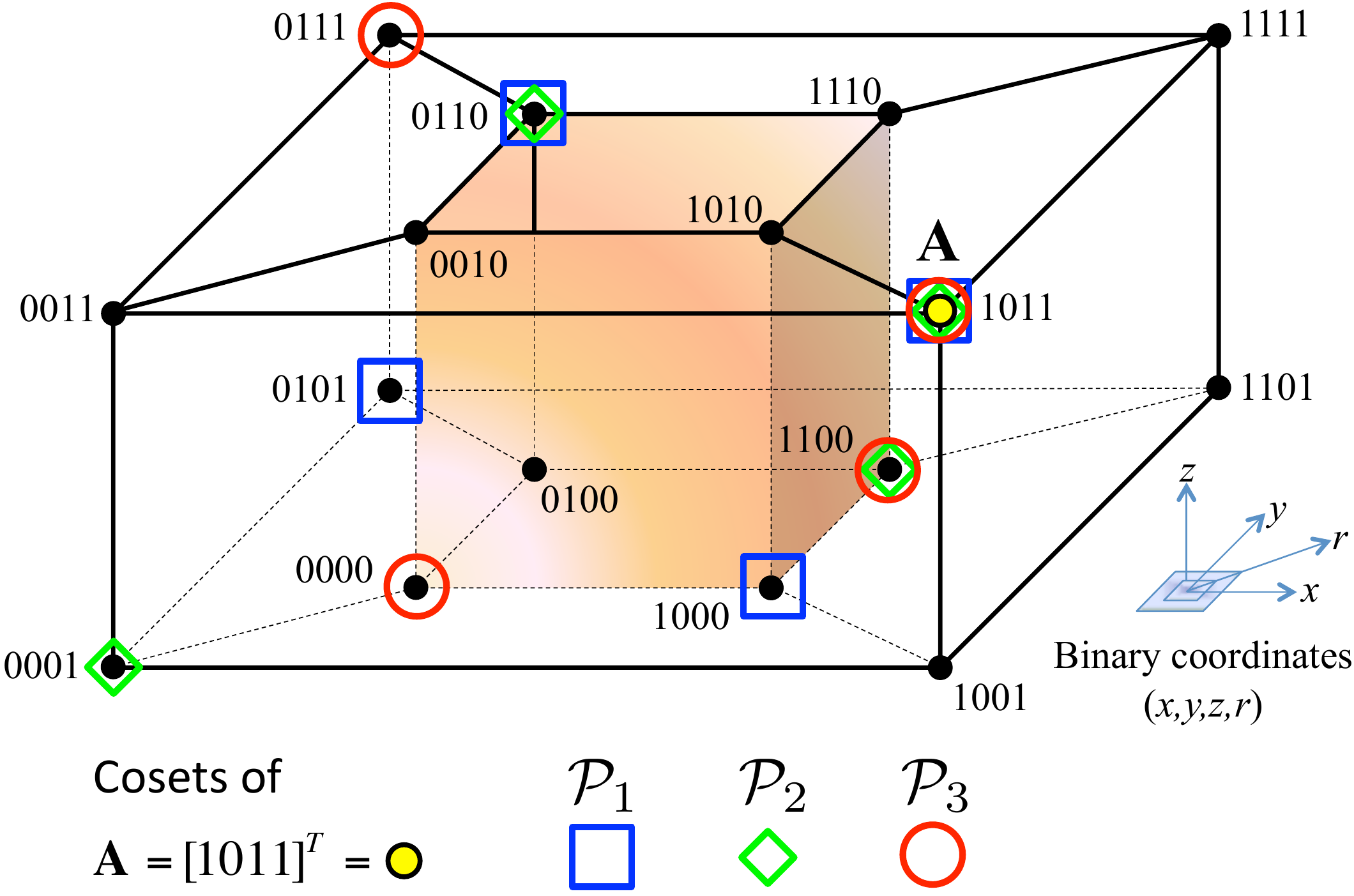}
\caption{This figure depicts the three codebook example of the sidebar to
  illustrate the design tension between maximizing security and
  privacy. This example concerns the space of 4-bit biometrics, which is illustrated by the $16$ points arranged on the vertices of a tesseract.
  The three cosets (with respect to each code) corresponding to enrollment biometric $\mathbf{A} = [1\,0\,1\,1]^T$ are depicted in this figure.}
\label{fig:multSystems}}
\end{figure}

Now, consider three additional secure sketch systems that use
$\mathbf{H}_1$, $\mathbf{H}_2$, $\mathbf{H}_3$, where $\mathbf{H}_1 =
\mathbf{H}$,
\begin{align*}
\mathbf{H}_2 = \left[\begin{array}{cccc} 1 & 0 & 1 & 1 \\ 0 & 1 & 0 & 1
    \end{array} \right], \ {\rm and} \ \,
\mathbf{H}_3 = \left[\begin{array}{cccc} 1 & 1 & 1 & 0 \\ 1 & 1 & 0 & 1
    \end{array} \right].
\end{align*}
For the same enrollment biometric $\mathbf{A} = [1011]^T$,
the syndromes, codewords and cosets are respectively given by 
$\mathbf{S}_1 = \mathbf{S}$, $\mathcal{C}_1 = \mathcal{C}$, and
$\mathcal{P}_1 = \mathcal{P}$; $\mathbf{S}_2 = [1 \, 1]^T$,
$\mathcal{C}_2 = \{[0000]^T, [1101]^T, [0111]^T, [1010]^T\}$, and
$\mathcal{P}_2 = \{[1011]^T, [0110]^T, [1100]^T,[0001]^T\}$;
$\mathbf{S}_3 = [0 \, 0]^T$ and $\mathcal{C}_3 = \mathcal{P}_3 =
\{[0000]^T, [1100]^T, [1011]^T, [0111]^T\}$. The geometry of
the cosets $\mathcal{P}_i$ is shown in Figure~\ref{fig:multSystems}.
There is linear
dependence between the codes defined by $\mathbf{H}_1$ and
$\mathbf{H}_2$ because of the shared first row and we observe
$|\mathcal{C}_1 \cap \mathcal{C}_2| = 2 > 1$. In contrast, the rows of
$\mathbf{H}_1$ and $\mathbf{H}_3$ are linearly independent and
$|\mathcal{C}_1 \cap \mathcal{C}_3| = 1$ due to only one intersection at the
origin, $[0000]^T$.  As we discuss next, linear independence between
the parity check matrices makes the systems more
secure, i.e., it reduces the SAR, but increases the potential for privacy
leakage.

First consider the SAR.  Say that the original
system, encoded using $\mathbf{H}$, is compromised, i.e., an attacker has 
learned the stored data $\mathbf{S}$.  Note that the attacker can gain access to System
1 with probability one, or SAR = 1.  This follows because
$\mathbf{H}_1 = \mathbf{H}$.  With knowledge of $\mathbf{S}$, the
attacker knows $\mathcal{P}_1=\mathcal{P}$ and can gain access
by uniformly setting $\mathbf{D}$  to be any member of
$\mathcal{P}$.  Recall that access is granted only if $\mathbf{D} \in \mathcal{P}$ 
since we have set $\tau = 0$.  If, instead of System 1, the attacker wants to access
System 2 using the same attack, i.e., by uniformly setting $\mathbf{D}$  to be any member of
$\mathcal{P}$. In this case, SAR $ = |\mathcal{P}
\cap \mathcal{P}_2| / |\mathcal{P}_2| = 0.5$. Finally, if the attacker wants to access
System 3, using the same strategy will result in an even smaller SAR of 0.25.  Note that
$0.25$ is also the nominal FAR for System 3, and so the attacker
does no better than random guessing.  The decrease in SAR is due to
the decrease in linear dependence between the parity check matrices:
reduced dependence implies reduced overlap in the respective cosets.

Next consider the privacy leakage.  Compromising the original system
meant that the attacker has discovered $\mathbf{S}$, thus 2 out of the 4 bits of 
$\mathbf{A}$ have been leaked. Suppose that, in addition to the original 
compromised system, the attacker could pick one more system to compromise.
Which system should he choose to obtain the most additional information about $\mathbf{A}$? 
Observe that  if the $i^{th}$ system is chosen for compromise, 
then $\mathbf{A} \in \mathcal{P} \cap \mathcal{P}_i$, so he wants the intersection
set to be as small as possible.  He learns nothing more about $\mathbf{A}$ 
by compromising System 1 since $|\mathcal{P} \cap
\mathcal{P}_1 | = | \mathcal{P}| = 4$.  However, as shown in Figure~\ref{fig:multSystems}, by 
choosing to compromise System 2 instead, his uncertainty of discovering $\mathbf{A}$ is 
reduced by one bit because
$|\mathcal{P} \cap \mathcal{P}_2 | = 2$.  Even better, by choosing to compromise
System 3, his uncertainty is completely eliminated and he discovers $\mathbf{A}$ because
$|\mathcal{P} \cap \mathcal{P}_3 | = |\{\mathbf{A}\} | = 1$.  Thus, the attacker would 
benefit the most by compromising the system with the {\em most} linear independence 
in its parity check matrix.


\centerline{\bf End Sidebar Inset \# B}

Using secure sketch for the purpose of illustration, this example 
shows how linearly independent parity check matrices make 
the systems most resistant to attack, but also most susceptible 
to privacy leakage. For simplicity, our example assumed identical
enrollment vectors $\mathbf{A}$ for all systems and a strict
threshold $\tau = 0$;  the tension between privacy leakage and SAR
also exists when the enrollment vectors used by the different
systems are noisy versions of each other and when the threshold $\tau$
is set to some nonzero value~\cite{wang12tifs,simoens09oakland}.

Analysis of linkage attacks can be further complicated when the
systems involved use different architectures.  For example, one system
may use secure sketch, another fuzzy commitment, and a
third cancelable biometrics.  Even when all systems have the same
architecture, it is still a difficult problem to select biometric
encoding parameters on each device to achieve a desired tradeoff
between security and privacy. In the analysis of~\cite{lai11tifs1,lai11tifs2}, 
information theoretically achievable outer bounds are derived for the privacy-security 
region for multiple systems. However, practical code designs that achieve these
bounds remain elusive.

It is also natural to ask what advantages and disadvantages result
when,  in the context of multiple systems, two-factor variants of the 
biometric architectures are used.  For secure sketches and fuzzy
commitments, each device can generate a different key then assigned to
the user. If the key or the stored data --- but not both --- is
compromised, there is no privacy leakage; the enrollment can be
revoked and new keys and new stored data can be assigned. However, a
(pessimistic) information theoretic argument shows that the SAR still
saturates to one whenever the stored data is
compromised, since an unbounded adversary could always
find an acceptable probe biometric (and key) through exhaustive search~\cite{wang12tifs}.
In the case of secure multiparty computation-based systems, the architecture extends in a
straightforward way to multiple systems: the user can simply choose a
different public-private key pair at each device. As long as
computational privacy guarantees hold, this strategy ensures that the
SAR remains low for devices whose decryption keys are not
compromised. However, if even one of the private keys is revealed, the
privacy leakage could be significant.  This is because the adversary
can access unencrypted information about the user's biometric feature
vector during the private distance computation protocol or during the
private comparison protocol. In the case of cancelable biometrics, the
user may employ a different non-invertible transformation at each
device, thereby ensuring that the SAR remains low for devices whose
specific transformation or stored data are not compromised.  However,
as noted earlier, the privacy leakage in the case of a compromised
transformation could be significant.

\section{Summary and Research Directions}

In this article, we have presented the main concepts that underlie
secure biometric systems and have described the principal
architectures by casting them as realizations of a single, general
authentication framework. Our objectives have been, first, to acquaint 
readers with the differences between secure and traditional biometric
authentication; next to familiarize them with the goals, ideas and
performance metrics common to all realizations of secure biometrics;
and finally to introduce them to the ways in which the various
realizations differ.  Table~\ref{fig:secbio-comparison} provides a
high-level summary of the secure biometrics architectures discussed,
comparing and contrasting their security assumptions, tools, complexity,
salient features and open problems.  

The study of secure biometric systems is a topical and fertile
research area.  Recent advances have addressed many aspects of secure
biometrics including new information theoretic analyses, the emergence
of new biometric modalities, the implementations of new feature
extraction schemes, and the construction of fast, encrypted-domain
protocols for biometric matching. That said, much work remains to be
done before secure biometric access control becomes commonplace.  
We now describe some research directions.
\begin{table*}[t]
\centering
\includegraphics[width=6.5in]{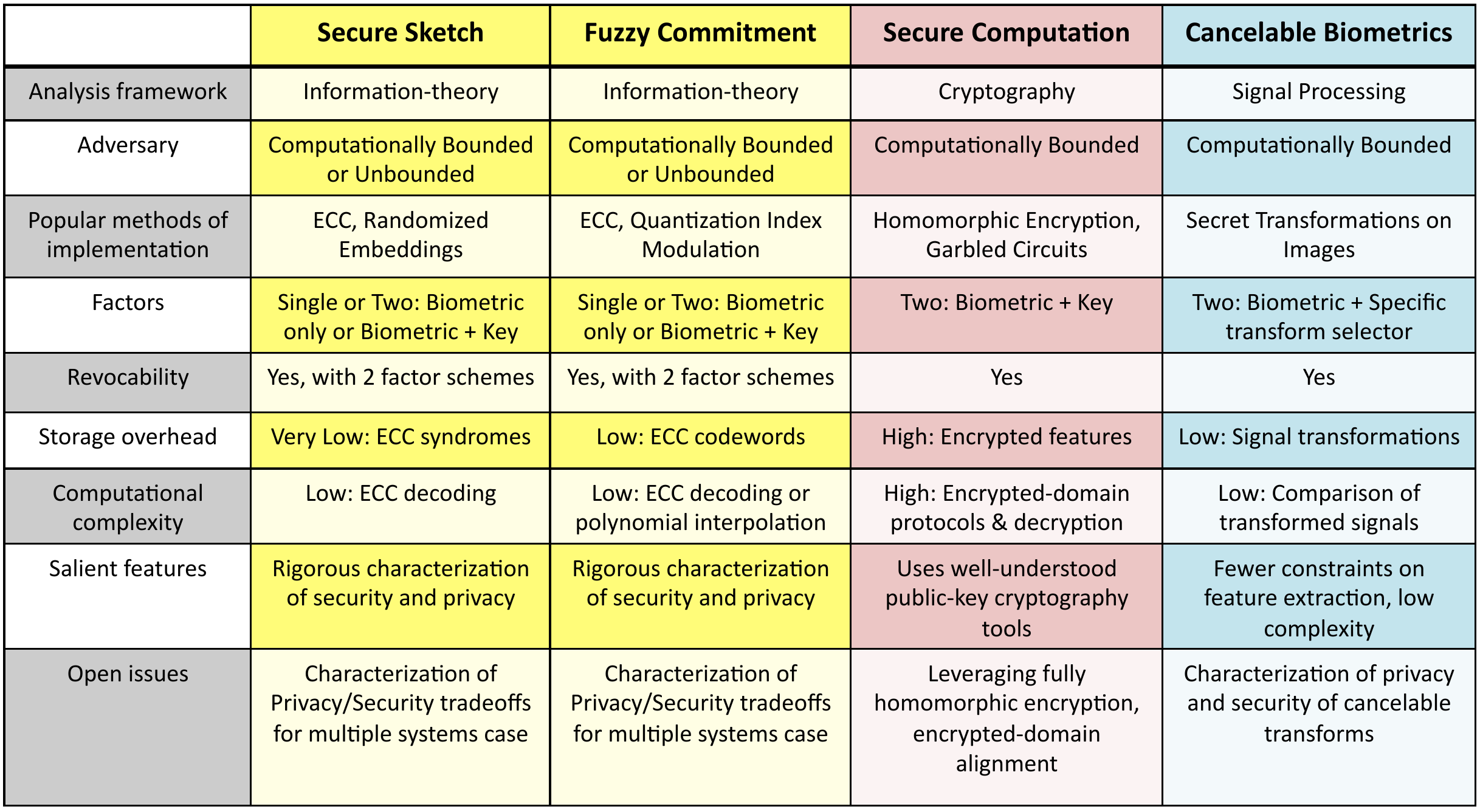}
\caption{A high-level comparison of the four secure biometrics architectures. \label{fig:secbio-comparison}}
\end{table*}
\\ \\ \textbf{Biometric Feature Spaces:} In almost all  secure
biometric system implementations, a traditional biometric feature
extraction technique is modified to make it compatible with one of the
privacy architectures that we have covered.  As observed in the
article, the incorporation of a ``secure'' aspect to the biometric
authentication system impacts the underlying tradeoff between FAR and
FRR.  The price of privacy is most often some drop in authentication
performance. The development of biometric feature spaces that provide
excellent discriminative properties, while simultaneously enabling
efficient privacy-preserving implementations, is the among the most
important current problems in the area. This is especially important
when the aim is to preserve discriminative properties in the context
of multiple systems in simultaneous use. As discussed in the article,
the effort involved would not be limited to signal processing
algorithms, but would also require evaluation of biometric
architecture using, for example, information theoretic or
game-theoretic problem formulations.  \\ \\ 
\textbf{Alignment and Pre-processing:} Much current work on implementation of secure
biometric systems ignores the fact that, prior to feature extraction
and matching in traditional biometric systems, a complicated and
usually non-linear procedure is necessary to align the probe and
enrollment biometrics. Traditional biometric schemes can store
alignment parameters such as shifts and scale factors in the clear.
But, for a secure biometric system, storing such data in the clear can
be a potential weakness.  On the other hand, incorrect alignment
drastically reduces the accuracy of biometric matching. Thus, it
is necessary to develop biometric matching schemes that are either
robust to misalignment, such as the spectral minutiae
method~\cite{xu09tifs}, or allow alignment to be performed under
privacy constraints.  \\ \\ \textbf{New Standardization Efforts:} In
addition to the development of novel approaches and methods,
widespread deployment of secure biometric systems will demand
interoperability across sensors, storage facilities, and computing
equipment.  It will also require an established methodology for
evaluating the performance of secure biometric systems according to
the metrics discussed herein. To this end, new standardization
activity has been undertaken in several domestic and international
bodies, composed of participants from industry, government and
academia~\cite{simoens12icb,iso24745}.  In the coming years,
standardization efforts are expected to address the task of
establishing guidelines and normative procedures for testing and
evaluation of various secure biometrics architectures.
\\ \\ \textbf{Attack Analysis and Prevention:} In this article, we have covered 
security attacks and privacy attacks, wherein the attacker attempts to extract 
information about the stored data, the biometric features and/or the keys 
and tries to gain unauthorized access to the system. In these attack
scenarios, the attacker does not disrupt or alter the system 
components themselves --- for example, change the ECC parity check matrix, 
thresholds, keys, or the biometric database, or arbitrarily deviate 
from the encrypted-domain protocols and so on. A comprehensive discussion 
of such attacks, including collusion with system administrators, and network-related attacks 
such as Denial-of-Service (DOS) attacks appears in~\cite{jain06tifs}. Modeling, 
experimental analysis and prevention of such attacks remains a very challenging topic
in academia and industry.
\\ \\ \textbf{Fully Homomorphic Encryption:} In the secure computation
community, much excitement has been generated by the discovery of
fully homomorphic encryption, which allows arbitrary polynomials to be
computed in the encrypted
domain~\cite{gentry10acm,van10eurocrypt}. Though current
implementations are exceedingly complex, faster and more efficient
constructions are emerging. These promise to be able, eventually, to
compute complicated functions of the enrollment and probe biometrics
--- not just distances --- using a simple protocol where nearly all
the computation can be securely out-sourced to a database server.
\\ \\ \textbf{Emerging Biometric Modalities:} Through the proliferation of
tablet computers, smartphones, and motion sensor devices for gaming,
many people have become familiar with touch and gesture-based
interfaces.  This has led to the emergence of new biometric
modalities. Authentication can be based on hand movements and
multi-touch gestures, leveraging techniques from machine learning and
computer vision~\cite{sae12hfcs,lai12avss}.  These modalities also
have an interesting property from the point of view of cancelability:
compromised templates can be revoked and renewed merely by having a
user choose a different gesture. Aspects of a gesture, such as body
shape and the relative sizes of limbs and fingers, generate features
that are irrevocable, just as with traditional biometrics. Incorporating
the dynamics of gestures into the authentication process has been shown
to improve FRR-FAR tradeoffs~\cite{lai12avss}.
In principle, there are an unlimited number of ways in which one
could personalize gestures that are reliable, easy to remember,
reproducible, and pleasant to work with. The study of gesture-based
biometric modalities is a nascent area of research.
\\ \\ \textbf{Related Applications:} As noted in the beginning of the article, many of the principles
discussed here extend to secure biometric identification systems. A practical concern is that
identification involves matching the test biometric against the entire database, which
means that the decision module in Figure~\ref{fig:highlevelBD} will be
executed once for each identity in the database. For large databases, ECC decoding or 
secure multiparty computation will be prohibitively complex unless fast, parallelizable 
algorithms are developed to compensate for the increased computational overhead. 
Other than authentication, these methods extend with minor modifications to the related
problem of secret key generation from biometrics. Furthermore, the concepts and methods
are readily applicable in emerging authentication scenarios that do not involve
human biometrics, e.g., device forensics and anti-counterfeiting technologies based on 
physical unclonable functions (PUFs)~\cite{holotyak11icassp,suh07dac,tuyls05fcds,skoric05acns}.


\bibliographystyle{IEEEtran}
\bibliography{references}

\section*{Author Biographies}

\noindent
\textbf{\textit{Shantanu Rane}} (Ph.D., Stanford University, 2007) is a Principal
Research Scientist at Mitsubishi Electric Research Laboratories in
Cambridge, MA. He is an Associate Editor of the IEEE Signal Processing
Letters,  and Transactions on Information Forensics and Security and is a member of the IFS
Technical Committee. He has 
participated in standardization activity for the
H.264/AVC standard, INCITS/M1 Biometrics, and the
ISO/SC37 Biometrics Subcommittee.
\\%
\\%
\textbf{\textit{Ye Wang}} (Ph.D., Boston University, 2011) is a Visiting Researcher
at Mitsubishi Electric Research Laboratories in Cambridge, MA. His research interests
include secure biometrics, information theoretically secure multiparty 
computation, and inference in networks.
\\%
\\%
\textbf{\textit{Stark C. Draper}}  (Ph.D., Massachusetts Institute of Technology, 2002) is an Assistant Professor at the University of Wisconsin, Madison.  He has held a research position at Mitsubishi Electric Research Labs (MERL) and postdoctoral positions at the University of California, Berkeley and the University of Toronto, Canada.  He has received the NSF CAREER award, the MERL 2010 President's Award, and a U.S. State Department Fulbright Fellowship.
\\%
\\%
\textbf{\textit{Prakash Ishwar}}  (Ph.D., University of Illinois at Urbana-Champaign, 2002) is an Associate Professor of Electrical 
and Computer Engineering at Boston University, an Associate Editor of the IEEE Transactions
on Signal Processing, and a member of the IEEE IVMSP Technical Committee. He was a recipient
of the 2005 US NSF CAREER award, a co-winner of the ICPR'10 Aerial View Activity Classification Challenge, and a co-recipient of the AVSS'10 best paper award.

\end{document}